\documentclass[DIV=calc, paper=letter, fontsize=11pt]{scrartcl}	 
\pdfoutput=1  

\usepackage[]{units}
\usepackage{lipsum} 
\usepackage[english]{babel} 
\usepackage[protrusion=true,expansion=true]{microtype} 
\usepackage{amsmath,amssymb,amsfonts,amsthm} 
\usepackage[svgnames, table]{xcolor} 
\usepackage[hang, small,labelfont=bf,up,textfont=it,up]{caption} 
\usepackage{booktabs} 
\usepackage{fix-cm}	 
\usepackage[]{units}

\usepackage{tabularx} 

\usepackage{hyperref}
\hypersetup{
colorlinks=true, 
linkcolor=blue, 
citecolor=blue, 
filecolor=blue, 
runcolor=blue, 
urlcolor = blue
}

\usepackage{fullpage}
\usepackage{sectsty} 

\usepackage{fancyhdr} 
\usepackage{lastpage} 

\usepackage{nicefrac}
\usepackage{cite}

\usepackage{tikz}
\usetikzlibrary{calc,patterns,decorations.pathmorphing,decorations.markings}
\usepackage{pgfplots}

\usepackage{multicol}
\usepackage{float} 

\newcounter{tempcolnum}
\makeatletter
\newcommand{\multicolinterrupt}[1]{
\setcounter{tempcolnum}{\col@number}
\end{multicols}
#1%
\begin{multicols}{\value{tempcolnum}}
}
\makeatother

\usepackage{graphicx}
\DeclareGraphicsExtensions{.png}

\usepackage{lettrine} 
\newcommand{\initial}[1]{ 
\lettrine[lines=3,lhang=0.3,nindent=0em]{
\color{DarkGoldenrod}
{\textsf{#1}}}{}}

\usepackage{caption}

\usepackage{float} 

\usepackage{graphicx}

\usepackage{titling} 

\newcommand{\HorRule}{\color{DarkGoldenrod} \rule{\linewidth}{1.0pt}} 

\pretitle{\vspace{-90pt} \begin{flushleft} \HorRule \fontsize{18}{18} \color{DarkRed} \selectfont} 

\title{\LARGE Electrostatics of Colloidal Particles Confined in Nanochannels: \\ Role of Double-Layer Interactions and Ion-Ion Correlations}

\posttitle{\par\end{flushleft}} 
\preauthor{\vspace{-10pt} \begin{flushleft}\fontsize{16}{16} \large  \color{DarkRed}} 
\author{Inderbir S. Sidhu$^*$ and Amalie L. Frischknecht$^{+}$ and Paul J. Atzberger$^{*}$ \hspace{6cm} } 
\postauthor{\fontsize{12}{12} \small \color{Black} 
$^{*}$ Department of Mathematics and Department of Mechanical Engineering, University of California Santa Barbara, 
e-mail: atzberg@gmail.com; website: \url{http://atzberger.org/}  \\
$^+$ Center for Integrated Nanotechnologies (CINT), Sandia National Laboratories. \\ 
\vskip 0.0em
\lfoot{\footnotesize * Work supported by DOE ASCR CM4 DE-SC0009254, W.M. Keck Foundation, \\ NSF CAREER Grant DMS-0956210, and NSF DMS - 1616353.}
\end{flushleft} \vspace{-17pt} \HorRule} 
\date{} 

\DeclareOldFontCommand{\bf}{\normalfont\bfseries}{\mathbf}

\newcommand{\mb}[1]{\mathbf{#1}}

\newcommand{\subtxt}[1]{ {\mbox{\tiny #1}} }

\definecolor{issuePJA_color}{rgb}{1.0,0.0,0.0}

\definecolor{commentPJA_color}{rgb}{1.0,0.0,0.8}

\definecolor{commentJKS_color}{rgb}{1.0,0.0,0.8}

\definecolor{commentALF_color}{rgb}{1.0,0.0,0.0}

\definecolor{AtzGrey}{rgb}{0.75,0.75,0.75}

\DeclareGraphicsExtensions{.png}

\begin{document}

\maketitle 
\thispagestyle{fancy} 


\vspace{-32pt}
\initial{W}\textbf{e perform computational investigations of electrolyte-mediated interactions of charged colloidal particles confined within nanochannels.  We investigate the role of discrete ion effects, valence, and electrolyte strength on colloid-wall interactions.  We find for some of the multivalent charge regimes that the like-charged colloids and walls can have attractive interactions.  We study in detail these interactions and the free energy profile for the colloid-wall separation.  We find there are energy barriers and energy minima giving preferred colloid locations in the channel near the center and at a distance near to but separated from the channel walls.  We characterize contributions from surface overcharging, condensed layers, and overlap of ion double-layers.  We perform our investigations using Coarse-Grained Brownian Dynamics simulations (BD), classical Density Functional Theory (cDFT), and mean-field Poisson-Boltzmann Theory (PB).  We discuss the implications of our results for phenomena in nanoscale devices.
}


\begin{multicols}{2}

\section{Introduction}
\label{sec:introduction}
In many microscale and nanoscale systems, electrolytes play a central role in collective interactions, equilibrium phase behaviors, and kinetics~\cite{SquiresQuakeFluidicsReview2005,
KirbyBook2010,BazantBookChapter2011}.  This includes transitions in the stability of colloidal suspensions~\cite{DerjaguinLandauColloids1941,OverbeekColloids1948,Hansen2000}, electrophoretic separation and detection in fluidic devices~\cite{PennathurTransport2004, SquiresQuakeFluidicsReview2005,BazantBookChapter2011,
KirbyBook2010,KirbyZetaPotentialReview2004}, and biomolecular interactions~\cite{BakerPNASElectrostatics2001,McCammonBoLiSCPF2015,McCammon1987}.  
Confinement of electrolytes and charged objects between charged walls presents additional effects often resulting in rich phenomena that are particularly important in nanoscale devices~\cite{SquiresQuakeFluidicsReview2005,
PennathurTransport2004}.  This owes in part to such features as the thickness of ionic layers becoming comparable to other length-scales in the system\cite{BaldessariDLOverlap2008,
KirbyBook2010,DasDLOverlap2010,
RobbinsMultiscaleElectroOsmosis2016}. 

For sufficiently charged multivalent systems additional phenomena can arise as observed in experiments and predicted by theory~\cite{PegadoLikeChargeAttraction2008,
GrierLikeChargeAttraction1997,
NetzSimilarChargedPlates2001,PincusTwoPlates2009}.  This includes the formation of condensed ion layers on surfaces, over-charging of walls and particles, and attractions between like-charged objects~\cite{GrierLikeChargeAttraction1997,
PincusChargeFluct2002,SaderAttractionUnresolved1999}.  
These effects have formed the basis for understanding phenomena such as DNA condensation~\cite{SafinyaDNACondensation2000, StevensDNACond2001,
PincusCondensationPolyelectrolytes1998,
Kuron2015,BloomfieldDNACondensation1991}, colloidal stability~\cite{NagornyakLikeChargeExp2009,
Hansen2000,GrierLikeChargeAttraction1997},
and attraction of like-charged plates~\cite{PegadoLikeChargeAttraction2008,
NetzSimilarChargedPlates2001,
PincusTwoPlates2009}.  

We further explore here phenomena of charged systems in the context of colloidal particles confined within nanochannels.  We investigate the behaviors of confined electrolytes and charged particles through coarse-grained molecular-level simulations using Brownian Dynamics (BD) and classical Density Functional Theory (cDFT).  We also make comparisons with predictions from mean-field Poisson-Boltzmann theory (PB).  We investigate the interactions between a charged colloidal particle and the nanochannel wall as the electrolyte concentration and particle charge are varied.  

We find that in some charge regimes the free energy of the particle as a function of its position within the channel develops significant minima in preferred locations near the channel center and near to but separated from the channel wall.  In some regimes these preferred locations are separated by significant energy barriers.  Motivated by nanofludic devices our results indicate that colloidal particles could exhibit interesting bi-modalities switching from long dwell-times in locations near the channel center to locations near the channel wall.  For instance, this could have implications for experimental protocols and devices such as capillary electrophoresis used in fluidics for separations and detection~\cite{wanDLOverlap1997,KirbyBook2010,
SquiresQuakeFluidicsReview2005,PennathurTransport2004}.

We investigate the origins of the free energy profile by using BD simulations to characterize at the coarse-grained molecular-level the ion-ion correlations and the surface overcharging and condensed ionic layers that form near the colloid surface and channel wall.  We further make comparisons with results from classical Density Functional Theory (cDFT).  We find the cDFT make predictions consistent with our molecular-level results but in the most strongly charged regimes with significant underestimation of the strength of effects such as the free energy well-depth.  For the free energy profile of the confined particle, the combined simulation and cDFT results demonstrate the significant roles played by ion-ion correlations and over-charging at both the charged walls and colloid particle surface.  We also show for the strongly charged regimes considered that a mean-field theory such as Poisson-Boltzmann theory is not adequate in predicting system behaviors highlighting the importance of accounting for ion-ion correlations and other discrete effects.

We introduce our BD simulations for the electrolyte and colloidal particle in Section~\ref{sec:rpm_model}.  We introduce our cDFT description of the nanochannel system in Section~\ref{sec:cDFT}.  We present the results of our calculations including the counterion and coion densities, colloidal particle free energy, and ion-ion correlation functions in Section~\ref{sec:results}.  We discuss our findings and related phenomena observed within nanochannels in Section~\ref{sec:discussion}.  Additional information on the computational methods developed and simulation protocols are discussed in Appendix~\ref{sec:detailsDFT} -~\ref{sec:corr_analysis}.

\lfoot{} 

\section{Electrostatics of the Nanochannel System}

\subsection{Brownian dynamics simulations}

\label{sec:rpm_model}

\begin{figure}[H]
\centering
\includegraphics[width=1.0\columnwidth]{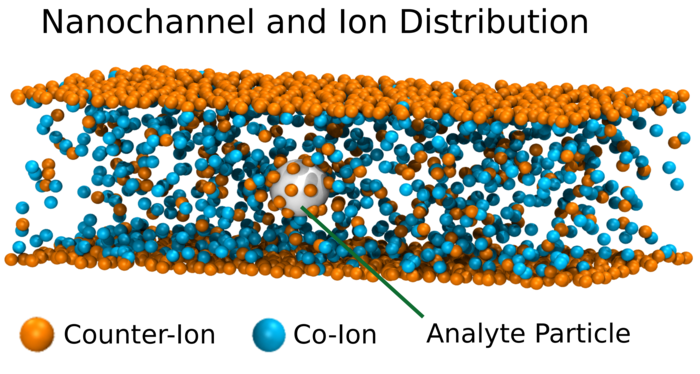}
\caption{Shown is a cut-away view of the electrolyte and colloidal particle corresponding to $\sigma = -6$ and $C_m = 8$ with counterions with $+2$ charge (orange) and coions with $-1$ charge (blue).  Strong correlations are exhibited, where counterions and coions form clusters and chains throughout the electrolyte and condensed layers near the walls and colloidal particle surface. For clarity the channel walls are not shown.}
\label{fig:rpm_model}	
\end{figure}

We consider colloidal particles confined within a nanochannel having a slit-like geometry. The walls of the channel are viewed as two like-charged parallel plates.  We consider electrolytes consisting of both counterions and coions, using a coarse-grained model related to the Restricted Primitive Model (RPM)~\cite{TorrieRPM_MC_1979,ValleauRPMElectrolytesII1980,
ValleauRPM_ElectrolytesI1980}. The discrete ion-ion interactions are taken into account within a continuous dielectric medium.  A snapshot of the system is shown in Fig.\ \ref{fig:rpm_model}.  After discussing our model for the ions, we discuss some additional details on the electrostatics of channels in Section~\ref{sec:electro_channels}.

We model the finite size of the ions and the excluded volume of the colloidal particle using the Weeks-Chandler-Andersen (WCA) interaction potential~\cite{WeeksChandlerAndersen1971}
\begin{eqnarray}
\nonumber
\label{whatever}
(\theequation)
\refstepcounter{equation}
\\
\nonumber
\phi_{\subtxt{wca}}(r)
= \left\{
\begin{array}{ll}
4\epsilon 
\left[
\left(
{b}/{r}
\right)^{12}
-
\left(
{b}/{r}
\right)^6
+
\frac{1}{4}
\right], & r \leq r_c \\
0, & r > r_c.
\end{array}
\right. .
\end{eqnarray}
The $r$ is the distance between the center-of-mass of the two particles.  For a particle with steric radius $b$ we have $r_c = 2^{1/6}\cdot b$.  This ensures a purely repulsive interaction between particles~\cite{WeeksChandlerAndersen1971}.  For the steric particle-wall interactions, we treat the walls as a smooth continuum and use the Lennard-Jones $9$-$3$ potential
\begin{eqnarray}
\label{eqn:phi_lj93}
\phi_{\subtxt{lj93}}(r)
= 
\epsilon 
\left[
\frac{2}{15}
\left(
{b}/{r}
\right)^{9}
-
\left(
{b}/{r}
\right)^3
\right].
\end{eqnarray}
Here, $r$ denotes the nearest distance between a particle and the wall.  Electrostatic interactions between ions and/or the colloidal particle of charge $q_1$ and $q_2$ are given by the Coulomb interaction
\begin{eqnarray}
\phi_{\subtxt{coul}}(r) = \frac{q_1q_2}{4 \pi \epsilon_0 \epsilon r}
\end{eqnarray}
where $\epsilon$ is the dielectric constant of the background medium and we use SI units. To account for the surface charge density $\sigma$ of the colloidal particle we use Gauss' Law~\cite{GriffithsBookEM1998}, allowing us to use a point charge $Q_0$ at the center-of-mass with $Q_0 = 4\pi R^2\sigma$, where $R$ is the radius of the particle.  

To handle the long range Coulumb interactions we use the Particle-Particle Particle-Mesh (PPPM) approach~\cite{Hockney1989,PollockPPPM1996} as implemented in LAMMPS~\cite{PlimptonLAMMPS1995}.  For the nanochannel with slit geometry we use a variant of the PPPM method which uses periodic boundary conditions in the xy-directions \cite{Yeh:1999dm}.  This method has been extended to allow the simulated system to have a net charge within the slab interior~\cite{Ballenegger:2009ct} which we utilize in our simulations.  Our overall system is electrically neutral with the electrostatics of channels with charged walls handled using our approach discussed in Section~\ref{sec:electro_channels}.

In some of the simulations, we use a harmonic potential to hold the colloidal particle at a given location by
\begin{eqnarray}
\label{eqn:phi_target}
\phi_{\subtxt{target}}\left(\mb{x}\right) = \frac{k}{2} \left|\mb{x} - \mb{x}_0 \right|^2, 
\end{eqnarray}
where $\mb{x}_0$ is the target location for the colloidal particle location $\mb{x}$. The total potential energy associated with a configuration of the nanochannel system including the counterions, coions, and colloidal particle is given by
\begin{eqnarray}
\label{equ:totalEnergy}
\Phi[\mb{X}] = \Phi_{\subtxt{coul}}[\mb{X}] + \Phi_{\subtxt{sterics}}[\mb{X}] + \Phi_{\subtxt{target}}\left[\mb{X}\right],
\end{eqnarray}
where we represent the configuration of colloidal particle and ions by the composite vector $\mb{X} = \lbrack \mb{X}_{\subtxt{colloidal-particle}}, \mb{X}_{\subtxt{ions}} \rbrack^T$.  To sample equilibrium configurations we use Brownian Dynamics (BD) based on the Langevin equations~\cite{Gardiner1985}
\begin{eqnarray}
m \frac{d \mb{V}}{dt} = -\gamma \mb{V} - \nabla \Phi[\mb{X}] + \mb{F}_{thm},
\end{eqnarray}
where $d\mb{X}/dt = \mb{V}$ and $\left \langle \mb{F}_{thm}(s) \mb{F}_{thm}^T(t) \right \rangle = 2k_B{T} \gamma\delta(t - s)$.  For the time integration we use a stochastic Velocity-Verlet method implemented within LAMMPS ~\cite{PlimptonLAMMPS1995,AtzbergerLAMMPS2016}.  All BD simulations are performed in LAMMPS, with parameter values as given in Table~\ref{table:defaultParams}.  

Throughout this paper we use BD to probe only equilibrium properties of the system.  The BD simulations were equilibrated from random initial conditions over times long enough for the ions to diffuse at least two times across the diameter of the nanochannel. We then collected statistics on trajectories in which the ions diffused at least five times across the nanochannel diameter.  

\begin{table}[H]
\tiny
\centering
\begin{tabular}{|l|l|}
\hline
\rowcolor{LightGrey}
\textbf{Parameter}  & \textbf{Value} \\
\hline
$\ell_z$      nanochannel width (z) & 6 $nm$ \\
$\ell_x,\ell_y$ nanochannel length (x,y) & 18 $nm$ \\
$\sigma_w$ wall surface charge & -0.72 $e/nm^2$ \\
$b_w$  wall steric parameter lj$93$ & 0.5 $nm$ \\
$r_c^{[w]}$ wall cut-off parameter lj93 & 0.425 $nm$ \\
$\epsilon_w$  wall energy lj$93$ & 2.27e+7 $amu\hspace{0.07cm}nm^2/ns^2$ \\
$\sigma$    particle surface charge & -3 $e/nm^2$ \\
$R$         particle radius  & 0.75 $nm$ \\
$m_0$       particle mass &  6.20e+3 $amu$ \\ 
${T}$       temperature             & 300 $K$ \\
$k_B{T}$    thermal energy          & 2.50e+6 $amu\hspace{0.07cm} nm^2/ns^2$\\
$\rho$      solvent mass density    & 6.02$e${+2}\hspace{0.07cm}$amu/nm^3$\\
$\mu$       solvent viscosity       & 5.36e+5 amu/(nm$\cdot$ns)\\
$\epsilon_r$  solvent relative permittivity & 80.1 \\
$b_{-}$  counterion radius & 0.116 $nm$ \\
$b_{+}$  coion radius & 0.116 $nm$ \\
$q_{-}$ counterion charge & -1 $e$ \\
$q_{+}$ coion charge & +2 $e$ \\
$m_{-}$  counterion mass + solvantion & 2.3$e${+1}\hspace{0.07cm}$amu$ \\
$m_{+}$  coion mass + solvantion & 2.3$e${+1}\hspace{0.07cm}$amu$ \\
$\bar{c}_{-}$ reference ion concentration & 0.214$M$ \\
$\bar{c}_{+}$ reference ion concentration & 0.128$M$ \\
$r_c^{[w]}$ wall LJ cutoff & 0.425 $nm$\\
$r_c^{[c]}$ coulombic cutoff & 6 $nm$\\
$\Delta{t}$ Langevin timestep & 1.0$e${-5}\hspace{0.07cm}$ns$ \\
$\gamma$ Langevin drag & $6 \pi\mu R$ \\
$\tau_e$ Langevin equilibration time & $0.5 ns$ \\
\hline
\end{tabular}
\caption{Parameter values for the nanochannel model.  We use by default these values unless specified otherwise.}
\label{table:defaultParams}
\end{table}

\subsubsection{Electrostatics of Channels}
\label{sec:electro_channels}

For channels having a slit geometry consisting of two parallel walls, the electrostatics exhibit a few interesting features.  For channels of finite extent with wall edges immersed in a reservior, the wall surface charges generate the 
strongest electric fields near the edges in the reservior.  Through cancellations in the Coulombic interactions the wall charges do not generate significant net electric forces on the ions toward the middle region of the channel away from reservior edges.  As a result, in the idealized limit of two infinite walls having equal and uniform surface charge, the electric fields generated by the wall-charges exactly cancel throughout the channel interior.

This can be seen by considering a single wall with charge $\sigma$.  This contributes to the electric potential for the ion interactions as
\begin{eqnarray}
\label{equ:intChargedWalls}
\\
\nonumber
\phi_{\subtxt{coul-w}}\left(z\right) 
= \int \frac{q_1\sigma(\mb{r}')}{\epsilon |z\mb{e}_z - \mb{r}' |} dx dy, 
\end{eqnarray}
where $\mb{r}' = x\mb{e}_z + y\mb{e}_z$.  The $\mb{e}_i$ denotes the standard basis vector pointing in the $i^{th}$ coordinate direction. For a constant uniform surface charge $\sigma$ this can be integrated to obtain the equivalent potential
\begin{eqnarray}
\phi_{\subtxt{coul-w}}(z) = -(2\pi q_1 \sigma/\epsilon) z.
\end{eqnarray}
For two equally charged parallel walls of infinite extent the net electric field
has a Coulombic potential that is independent of $z$.  This can be seen from
\begin{eqnarray}
\nonumber
(\theequation) \hspace{1.4cm}
\label{equ:wallZeroField}
\refstepcounter{equation}
\\
\nonumber
\phi(z) &=& 
\phi_{\subtxt{coul-w}}\left(z\right)
+ \phi_{\subtxt{coul-w}}\left(L - z\right) \\
\nonumber
&=& -(2\pi q_1 \sigma/\epsilon) \left(z + L - z\right) \\ 
\nonumber
&=& -(2\pi q_1 \sigma/\epsilon) L.
\end{eqnarray}
As a consequence, the net electric field $E = -d\phi/dz$ acting on ions confined between the walls is zero.  

It is worth mentioning that such cancellations would not hold in the case of two walls that have a finite extent or non-uniform surface charge.  For equal uniform charges this can be seen by integrating equation~\ref{equ:intChargedWalls} in polar coordinates for two disk-like walls of radius $R$.  Our results show that for uniformly charged walls as their extent becomes large the electric fields contribute negligably toward the middle region of the channel away from the reserviors.

These results suggest a few interesting mechanisms by which ion concentrations are determined in the middle region of the channel and overall electric neutrality is acheived.  The results indicate that the electric fields generated by the walls near the reservior edges of the channel are primarily responsible for driving ions into the channel or expelling them to acheive electric neutrality.  Also, in the middle region of an infinite channel, the lack of net electric force acting on the ions from the walls gives an interesting perspective on the electric double-layers.  Rather than conceiving of ions being pulled toward the charged walls, our results indicate once ionic concentrations are setup from the edge effects, the double-layer structures should be viewed as arising from how the walls break symmetry.  In particualr, since like-charged ions repel one another within the confined region and there are no balancing forces from ion charges on the other side of the walls, the like-charged ion repulsions can be viewed as pushing ions from each other from the channel interior towards the walls.  This occurs in a manner very similar to mechanisms underlying generation of osmotic pressures~\cite{AtzbergerOsmosis2007,AtzbergerWuPeskinOsmosisVesicle2015}.  It is in this manner that the double layers can arise in the channel middle region without the need for local net electric forces generated by the two walls.  From electric neutrality the ion concentrations are determined and such double-layers can be related to the Poisson-Boltzmann theory (PB) for single and two charged walls.

Our simulations capture such phenomena in the middle region of charged channels.  We use periodic boundary conditions to capture behaviors similar to the limit of walls of infinite extent.  Since in this limit the walls exert no net electric force on the ions, we handle implicitly the contributions of the wall charge.  Our approach is similar to the Ewald summation method of Ballenegger et al \cite{Ballenegger:2009ct}.  In this approach the energy of the charged slab system is regularized by placing two charged walls above and below the simulation system, with charge densities that neutralize the system.  Thus, we are simulating a system that is overall electrically neutral with two walls of an appropriately chosen equal charge that serve to balance the ions. 

For mean-field Poisson-Boltzmann theory (PB), charged walls are often handled by employing Neumann boundary conditions to account for surface charge explicitly~\cite{XingWallsPB2011,Maduar2016,NetzInterfaces2016,Netz2000}.  A crucial consideration linking this to our molecular perspective is the condition of electric neutrality.  For channels this implies the implicit determination of a surface charge for the walls.  For our model, electric neutrality allows us to distinguish different choices for the wall charge which result in an excess or deficit of ionic species in the interior region driven by the edge electric fields.  In this manner our molecular model gives overall results that can be directly related to continuum models with explicit Neumann boundary conditions for the wall charge~\cite{Maduar2016,Netz2000}.  We discuss how the ionic species concentrations in the channel interior are related to the implicit choice of the wall charge in Section~\ref{sec_model_params}.

\subsubsection{Model parameters}
\label{sec_model_params}
We investigate the structure of the double-layer as the strength of charge of the colloidal particle and as the ion concentrations are varied. We characterize the charge of the negatively charged colloidal particle $Q_\subtxt{particle}$ in terms of its surface charge density $\sigma$, where $Q_\subtxt{particle} = 4\pi R^2 \sigma$.  We performed simulations for colloidal particles with surface charge densities of $\sigma = $ -1, -3, and -6 e/nm$^2$; for brevity we will refer to these three cases without units as the systems with $\sigma$ = -1, -3, and -6. We mostly focus on divalent cations with $q_+ = 2e$ and monovalent anions with $q_- = -1e$.  We take as a reference concentration for the counterions $\bar{c}_{+} = 0.128 M$ and for the coions $\bar{c}_{-} = 0.214 M$, expressed in molar units.  Other ion concentrations are a multiple $C_m$ of these baseline reference concentrations.  For example, $C_m = 10$ corresponds to a counterion concentration $c_{+} = C_m\bar{c}_{+} = 1.28 M$ and a coion concentration $c_{-} = C_m\bar{c}_{-} =  2.14 M$. The simulations are performed with a fixed number of ions, with an excess of counterions so that the bulk electrolyte solution is not neutral. The excess counterions (cations) lead to an effective negative charge on the nanochannel walls, given by the condition of overall electric neutrality:
\begin{eqnarray}
\label{eqn:neutral}
\\
\nonumber
q_{-}N_{-} + q_{+}N_{+} + Q_\subtxt{particle} + 2Q_\subtxt{wall} = 0.
\end{eqnarray}
Here $N_{-} = V c_{-}$ and $N_{+} = V c_{+}$ denote the number of ions in the unit cell where $V$ is the channel volume.   $Q_\subtxt{wall}$ is the charge on each wall in the unit cell.  For a given fixed concentration of coions and counterions the effective surface charge of the wall is obtained from electric neutrality by solving for $Q_\subtxt{wall}$ in equation~\ref{eqn:neutral}. The wall surface charge density for each system simulated is given in units of e/nm$^2$ in Table~\ref{table:wall_charge}.  The wall charge density increases with increasing ion concentration. Additionally, the wall surface charge densities vary slightly depending on the colloidal particle charge, since we have a fixed number of ions in the channel. 

\begin{table}[H]
\begin{center}
\begin{tabular}{|lccc|}
\hline
\cellcolor{AtzGrey} & \cellcolor{LightGrey} $\mathbf{\sigma = -1}$ &  \cellcolor{LightGrey}$\mathbf{\sigma = -3}$ & \cellcolor{LightGrey} $\mathbf{\sigma = -6}$\\
\cellcolor{LightGrey}
$\mathbf{C_{m} = 1}$ & -0.74 & -0.72 & -0.68 \\
\cellcolor{LightGrey}
$\mathbf{C_{m} = 2}$ & -1.49 & -1.46 & -1.43 \\
\cellcolor{LightGrey}
$\mathbf{C_{m} = 4}$ & -2.98 & -2.96 & -2.93 \\
\cellcolor{LightGrey}
$\mathbf{C_{m} = 6}$ & -4.48 & -4.46 & -4.43 \\
\cellcolor{LightGrey}
$\mathbf{C_{m} = 8}$ & -5.98 & -5.95 & -5.92 \\
\cellcolor{LightGrey}
$\mathbf{C_{m} = 10}$ & -7.47 & -7.45 & -7.42 \\ 
\hline
\end{tabular}
\caption{Wall Surface Charge Density (units are $e/nm^2$).  For the different regimes considered, we give the implicit surface charge density that arises from electric neutrality given by the condition in equation~\ref{eqn:neutral}. 
}
\label{table:wall_charge}
\end{center}
\end{table}

In the regimes we consider, the electrostatic interactions vary in strength. We can characterize the strength of the interactions by the electrostatic coupling constant \cite{NetzStrongCouplingTheory2000} given by
\begin{equation}
	g \equiv 2 \pi q^3 \ell_B^2 \sigma.
\end{equation}
Here $q=2e$ is the charge of the divalent counterions and $\sigma$ is the charge density of either the colloidal particle or the channel walls. The Bjerrum length $\ell_B$, the distance at which the electrostatic interaction energy is comparable to the thermal energy 
$k_B{T}$, is $\ell_B \equiv e^2/4 \pi kT \epsilon\epsilon_0$. In our systems with divalent cations, the electrostatic coupling constant ranges from $g \approx 17$ for the least charged system, up to $g \approx 188$ for the most strongly charged system. Previous studies of electrolytes near flat surfaces\cite{NetzStrongCouplingTheory2000} have shown that the counterion density profiles agree with the PB theory for $g \approx 1$, the profiles show clear deviation from PB theory for $g=10$ and $g=100$, and they show good agreement with the strong-coupling limit for $g=10^4$ see \cite{NetzStrongCouplingTheory2000}. Previous simulations of highly charged spheres explored coupling constants ranging from $g=26$ up to $g = 615$ and found attraction between like-charged spheres \cite{Allahyarov:1998kz,GronbechJensen:1998em,StevensFrischknechtNanoparticle2016}. We therefore expect our simulations to be in the intermediate regime between weak and strong coupling.

\subsection{Classical Density Functional Theory (cDFT)}
\label{sec:cDFT}

In the classical density functional theory (cDFT) calculations, we use the original form of the RPM, i.e. we model the ions as interacting charged hard spheres with diameters $d_\alpha$ and charges $q_\alpha$, in a background continuum dielectric medium to represent the solvent. We represent the colloidal particle as a larger hard sphere of radius $R$ that has surface charge density $\sigma$.  The ions are treated as mobile fluid species, while the colloidal particle has a fixed spatial location. We account for the steric interactions between the ions and the colloidal particle using a hard sphere interaction $V(r) = \infty$ for $r<R$, where $r$ is the distance between the ion and the center of the colloidal particle.  In addition, we add a smooth truncated potential based on the Lennard-Jones (LJ) interaction to the surface of the colloidal particle, 
\begin{equation}
V^{mLJ}_\alpha(r') = 4\epsilon_m \left[\left(\frac{\sigma_m}{r'}\right)^{12} - \left(\frac{\sigma_m}{r'}\right)^{6} \right],
\end{equation}
where $r'$ is the distance between the ion and the surface of the colloidal particle. We truncate and shift this potential to obtain 
\begin{equation}
V^{m}_\alpha(r') = V^{mLJ}_\alpha(r') - V^{mLJ}_\alpha(r'_c), \hspace{0.4cm} r' < r'_c,
\end{equation}
with $V^{m}_\alpha(r')=0$ for $r'>r'_c$, at large distances from the colloidal particle. In our notation, the subscript $\alpha$ refers to the index of the particular ion species and the $mLJ$ and $m$ to the modified Lennard-Jones potentials.  This repulsive potential serves to smooth the surface of the colloidal particle to reduce mesh-size effects in our discretized cDFT. We used $\epsilon_m = 0.5k_BT$ and $\sigma_m= d$ (where $d$ is the ion diameter) for all calculations. The channel boundaries are modeled as hard walls with the interaction potential for the ions 
\begin{equation}
	V^w_\alpha(z) = \left\{ \begin{array}{ll}
	\infty, &  \mbox{ions outside the channel} \\
	0,   &  \mbox{ions inside the channel}.
	\end{array} \right.
	\label{eq:HS}
\end{equation}
The volume of fluid trapped between the two channel walls is referred to as the ``inside" region and everything else as ``outside" of the channel.  This potential imposes that ions can not penetrate the walls and must remain within the channel region between the two walls. 

We use a formulation of cDFT that follows closely the work of Oleksy and Hansen \cite{Oleksy:2006ed} and is very similar to that of Henderson et al.\cite{Henderson:2011fn}.  We formulate the cDFT for an open ensemble, specified by the temperature $T$, the total volume $V$, and the chemical potentials $\mu_{\alpha}$ of all fluid species in the system.  We discuss the relation of these parameters to those used in the BD simulations in Section~\ref{subsec:cdft_param}.

The grand free energy of the system is given as a functional of the ion densities $\rho_{\alpha}(\bf r)$:
 \begin{eqnarray}
 	\label{eq:omega1}
	\Omega[\rho_\alpha({\bf r})]  = \sum_\alpha F[\rho_\alpha({\bf r})] \hspace{2cm} \\
	- \sum_\alpha \int d{\bf r} \left(\mu_\alpha - V_\alpha({\bf r})  \right) \rho_\alpha({\bf r}).
\end{eqnarray}
For notational convenience, it is to be understood that $\Omega[\rho_\alpha({\bf r})]$ depends on all of the density fields $\{\rho_\alpha\}$ collectively, where we use this convention to reduce clutter.  Here $F[\rho_\alpha({\bf r})] $ is the intrinsic Helmholtz free energy of the system.  $V_\alpha({\bf r}) = V+ V^m + V^w$ denotes the neutral part of the potential that acts on each ion from the walls and the colloidal particle.  The equilibrium density profile $\rho^0_\alpha({\bf r})$ minimizes the free energy functional $\Omega[\rho_\alpha({\bf r})]$.  This can be expressed in terms of the variational derivative~\cite{Gelfand2000} 
\begin{equation}
	\left. \frac{\delta \Omega[\rho_\alpha({\bf r})]}{\delta \rho_\alpha({\bf r})}  \right|_{\rho_\alpha^0} = 0.
	 \label{eq:domega}
\end{equation}
At equilibrium the associated grand potential free energy of the system is $\Omega^0 = \Omega[\rho^0_\alpha({\bf r})]$ \cite{Evans:1979jn}. The intrinsic Helmholtz free energy consists of four terms given by
\begin{eqnarray}
\label{eq:Fhelm}
F[\rho_\alpha({\bf r})]  & = & F_{id}[\rho_\alpha({\bf r})] + F_{hs}[\rho_\alpha({\bf r})] \\ 
\nonumber
& + & F_{coul}[\rho_\alpha({\bf r})]  + F_{corr}[\rho_\alpha({\bf r})].	
\end{eqnarray}
The terms represent respectively the Helmholtz free energies for the ideal gas (id), hard spheres (hs), mean-field Coulombic interactions (coul), and second order charge correlations (corr).  In formulating the DFT, approximations are needed to capture each of the listed effects.  We give more details in Appendix~\ref{sec:detailsDFT}.  

We emphasize the importance of the ion-ion correlation term $F_{corr}$ in cDFT which allows for capturing higher-order effects of density fluctuations distinguishing the cDFT results from those of mean-field theories like Poisson-Boltzmann (PB) theory.  As we shall show these correlations play an especially important role in the ion distributions observed in multivalent systems.  Without the correlation term (corr) and steric term for hard spheres (hs), the free energy functional $F$ reduces to that of the Poisson-Boltzmann theory.  By including or excluding the different terms in the free energy $F$ we can investigate different theories for the relative contributions of various effects on the observed ion distributions and colloid-wall interactions.  We now briefly discuss each of the terms in equation~\ref{eq:Fhelm}.

The term $F_{id}$ corresponds to the contributions of an ideal gas which for a given density is known exactly and is given in Appendix~\ref{sec:detailsDFT}.  For the hard-sphere interactions $F_{hs}$, we use the \textit{White Bear} version of the fundamental measure theory ~\cite{Roth:2002p518}.  The mean-field Coulombic interaction $F_{coul}$ is given by integrating the collective electric potential and density of the ionic species, see Appendix~\ref{sec:detailsDFT}.  The charge correlation term $F_{corr}$  is based on a functional Taylor expansion of the direct correlation function, which in turn is obtained from the known analytic solution of the mean-spherical approximation (MSA) for mixtures of charged hard spheres given in~\cite{Oleksy:2006ed}. Detailed expressions for each of these free energy terms are given in Appendix~\ref{sec:detailsDFT}. 

Minimization of the grand free energy in equation~\ref{eq:omega1} with respect to the density profiles of each ionic species is expressed mathematically as a set of nonlinear partial differential-integral Euler-Lagrange (EL) equations.  We express this in terms of residual equations $R_i = 0$ where
\begin{eqnarray}
	\label{eq:r1}
	R_1 & = & \ln \rho_\alpha({\bf r}) + V_\alpha({\bf r}) - \mu_\alpha \\
	\nonumber
	& + & \int \sum_\gamma \frac{\partial \Phi}{\partial n_\gamma} ({\bf r'}) \omega_{\alpha}^{(\gamma)}(\mathbf{r} - \mathbf{r}') d {\bf r'}  \\
		\nonumber
	& - & \sum_\beta \int d{\bf r'} \rho_\beta({\bf r'}) \Delta c_{\alpha\beta}({\bf r-r'}) + Z_\alpha \phi({\bf r})
			\nonumber \\			
			\nonumber
	\label{eq:r2}		
	\\	
	\nonumber
	R_2 & = & n_\gamma({\bf r}) - \sum_{\alpha}\int d\mathbf{r}' \,  \rho_{\alpha}(\mathbf{r})
                \omega_{\alpha}^{(\gamma)}(\mathbf{r} - \mathbf{r}') \\   
		\label{eq:r3}                             
                \\
	R_3 & = & \nabla^2 \phi({\bf r}) +\frac{4\pi \ell_B}{d} \sum_\alpha q_\alpha \rho_\alpha(\mathbf{r})
\end{eqnarray}
Here $\phi$ is the electric potential; other terms are defined in Appendix~\ref{sec:detailsDFT}.  The residual equations are solved computationally within the spatial domain of the nanochannel.  The third residual equation $R_3$ is Poisson's equation for the electrostatic potential $\phi({\bf r})$.  The cDFT calculations are performed using the open source package Tramonto, available at \url{https://github.com/Tramonto/Tramonto}.  The EL equations are solved in real-space on a Cartesian mesh using inexact Newton iterations for the density fields and a finite element method for the electrostatic potential.  Details of these numerical methods and discussions of related applications of Tramonto to charged systems can be found in~\cite{FrischknechtCDFTNumerical2002,
HerouxCDFTNumerical2007,BAM,Frink:2012hn}. 

All quantities in the residual equations have been expressed in terms of reduced units with energies in units of $k_BT$ and lengths in units of the ion diameter $d$.  $Z_\alpha$ is the valence of species $\alpha$.   The dimensionless quantity appearing in $R_3$ is sometimes called the plasma parameter or the reduced temperature, $T^* = d/\ell_B$. 

\subsubsection{Parameterization}
\label{subsec:cdft_param}
We parameterized our cDFT calculations to yield results in comparable physical regimes as the BD simulations.  This was done by taking the temperature and dielectric constant so that $\ell_B =$ 7.1 {\AA} as in the simulations, using the same surface charge density on the colloidal particle, matching the ion diameters $d = 2b =$ 0.232 nm, and using the radius $R = $ 0.75 nm for the colloidal particle.  We used a channel with total width $\ell_z$ = 6 nm as in the simulations. The channel walls extend into the channel to the same distance as in the simulations, so that we match the hard wall condition in the DFT with the Lennard-Jones 9-3 repulsive walls in the simulations.  

To reduce computational costs in the cDFT calculations, we placed the colloidal particle with its center on the z-axis, so that the symmetry of the system allows for reflecting boundary conditions to be used in the x- and y-directions and thus only 1/4 of the particle needs to be directly included in the calculations.  For this purpose, the size of the computational domain in the x and y directions was $\ell_x = \ell_y = $ 4.6 nm, for an effective channel length of 9.28 nm (taking into account the reflecting boundary through the center of the particle). We used a mesh size of 0.058 nm in all the 3D calculations (i.e. a mesh size of 0.25$d$ in reduced units, where $d$ = 0.232 nm is the diameter of the ions). 

The BD simulations were performed in the canonical ensemble at constant $N_\alpha$, $V$, and $T$.  For cDFT it is more natural to work in the grand canonical ensemble at constant $\mu_\alpha$, $V$, and $T$.  To make a correspondence between these two sets of calculations, we set the chemical potentials in the cDFT so that the average ion densities match the BD simulations at the middle of the channel where nearly bulk conditions prevail. In the middle of the channel, the electrolyte solution is neutral, with $c_- = 2c_+$.  We set the surface charge density of the channel walls in the cDFT equal to the effective surface charge densities given in Table \ref{table:wall_charge}. 

We solve equations (\ref{eq:r1})-(\ref{eq:r3}) in the nanochannel geometry with Neumann boundary conditions on $\phi({\bf r})$ at the nanochannel walls and the colloidal particle, i.e.\ we set the charge density of these surfaces. We employ Dirichlet boundary conditions elsewhere, with a reflecting boundary through the colloidal particle as described above.

To obtain the free energy associated with the particle at a particular position within the channel, we performed a cDFT calculation at each particle position and use the grand free energy of the resulting density.  We computed density profiles of ions around the particle both in the case with the particle in the center of the channel and in the case with the particle in the bulk fluid with no channel present.  The density profiles were found to be the same in both cases.  We also found that the density profile near the channel wall, at a location in the channel far from the particle, was also independent of the presence or absence of the colloidal particle.  This allowed us a significant reduction in computational costs by performing calculations of the wall density profiles from 1D systems using cDFT.  In our 1D calculations we used a finer mesh size of 0.0232 nm for better resolution in the reported results. 

\subsection{Poisson-Boltzmann (PB): Mean-Field Theory}
\label{subsec:PB}
In the limit that the ions are treated as point particles and do not have any charge correlation contribution to their free energy, the cDFT reduces to the Poisson-Boltzmann (PB) equation.  The PB limit corresponds to the  Helmholtz free energy functional with only the ideal gas and mean-field Coulombic contributions given by
\begin{eqnarray}
\label{equ:pb_free_energy}
\\
\nonumber
	 \beta F[\rho_\alpha({\bf r})] &  = &  \beta F_{id}[\rho_\alpha({\bf r})] +  \beta F_{coul}[\rho_\alpha({\bf r})]  \\
	 \nonumber
 & = & \sum_\alpha  \int d{\bf r} \rho_\alpha({\bf r}) \left(\ln \rho_\alpha({\bf r}) - 1\right)  \\
 	 \nonumber
 & + & \sum_\alpha \int d{\bf r}  q_\alpha \rho_\alpha({\bf r}) \phi({\bf r}),
\end{eqnarray}
where $\beta = 1/kT$.  Minimization of the grand free energy in equation~\ref{eq:omega1} using the free energy $F$ in equation~\ref{equ:pb_free_energy} gives
\begin{eqnarray}
	\frac{\delta \Omega}{\delta \rho_\alpha} = 0 = \ln \rho_\alpha({\bf r}) +  q_\alpha \phi({\bf r}) - \beta \tilde{\mu}_\alpha.
\end{eqnarray}
Here $\tilde{\mu}_\alpha = \mu_\alpha - V_\alpha(\mb{r})$ is the spatially dependent chemical potential including the contributions of the ion interactions with the channel wall and colloidal particle in equation~\ref{eq:omega1}.  Solving for the density gives
\begin{equation}
\label{equ:pb_density}
	\rho_\alpha({\bf r}) = \exp [\beta \tilde{\mu}_\alpha - q_\alpha \phi({\bf r})].
\end{equation}
In the case that the electric potential vanishes to zero in the bulk we have $\rho_\alpha^b = \exp[\beta \tilde{\mu}_\alpha]$.  However, in the nanochannel system  the term $\rho_\alpha^b$ should be interpreted with some care.  Since the steric interaction potential depends on ion location we technically have $\rho^b_\alpha(\mb{r}) = \exp[\beta \tilde{\mu}_\alpha(\mb{r})]$, which is a known function of position.  However, in the limit of hard wall interactions that we use here, the PB theory can be further simplified by using boundary conditions to represent the walls and colloidal particle.  This eliminates the explicit dependence of 
$\rho^b_\alpha(\mb{r})$ on position.  The remaining part of the chemical potential $\mu_\alpha$ is constant and we simply have $\rho_\alpha^b = \exp[\beta \mu_\alpha]$, where $\rho_\alpha^b$ are the reference densities (ion densities in a reservoir in equilibrium with the nanochannel system; these are nearly identical to the ion densities in the middle of the channel).

The electric potential satisfies Poisson's equation $\nabla^2 \phi = -(4\pi \ell_B/d) \sum_\alpha \rho_\alpha$.  Combining this with the densities found in equation~\ref{equ:pb_density} gives the non-linear Poisson-Boltzmann equations
\begin{equation}
\nabla^2 \phi({\bf r}) = -\frac{4\pi \ell_B}{d} \sum_\alpha \rho^b_\alpha \exp[- q_\alpha \phi({\bf r})].
\end{equation}
Here $d$ is a reference length in the system which for convenience we take to correspond to the ion size but other choices are also possible.

\section{Results}
\label{sec:results}

We first discuss results of the BD simulations, followed by comparisons with cDFT and PB theory. All figures show results from the BD simulations unless explicitly noted otherwise.

\subsection{Ionic Double-Layer Structure: BD Simulations}

\label{sec:ion_dl_struct}
\begin{figure}[H]
\centering
\includegraphics[width=0.99\columnwidth]{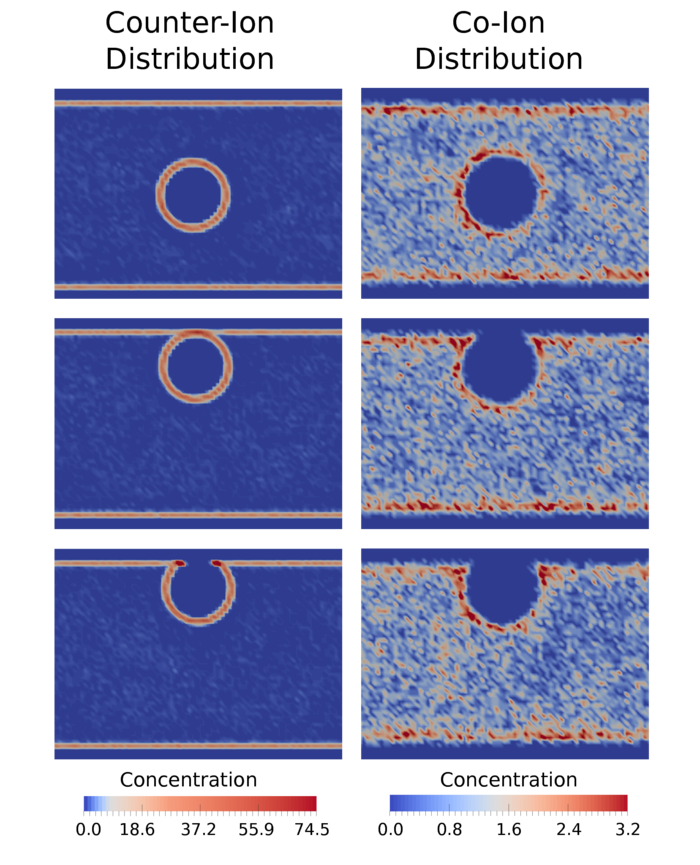}\caption{The average concentration of counterions (left) and coions (right) as the colloidal particle position is varied within the nanochannel, at $X_0^{(3)}$ = 3.0 nm, 4.6 nm, and 4.85 nm (top to bottom), for $\sigma = -3$ and $C_m = 8$.}
\label{fig:avgDensity2D}	
\end{figure}

We show in Figure~\ref{fig:avgDensity2D} typical distributions for the counterions and coions as the colloidal particle position is varied in the case of $\sigma = -6$ and $C_m = 8$.  In this regime strong layering occurs for the counterions near the walls and near the colloidal particle surface.  Also, a secondary layer of coions occurs offset from the walls and the colloidal particle surface adjacent to the counterion layer. This is especially visible for the coions shown in the right panel of Figure~\ref{fig:avgDensity2D}.  

\begin{figure}[H]
\centering
\includegraphics[width=0.99\columnwidth]{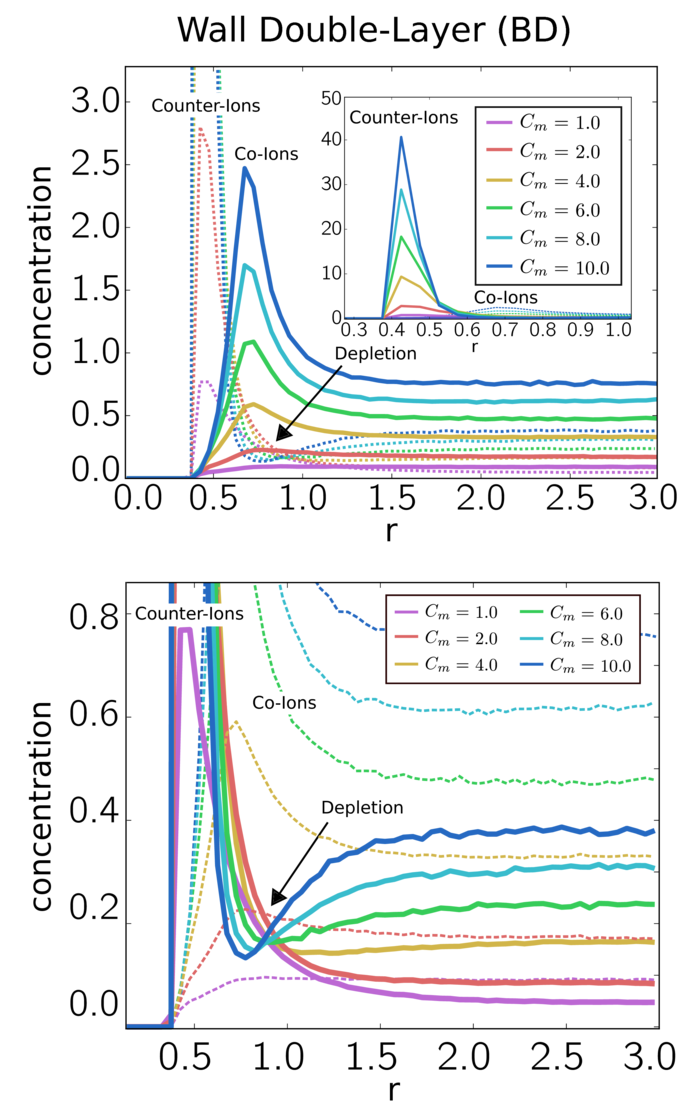}\caption{Ion concentrations near the channel walls as $C_m$ is varied, for $\sigma = -6$.  }
\label{fig:wall_layers}	
\end{figure}

We show the ion concentrations near the wall for $\sigma = -6$ and varying $C_m$ in Figure~\ref{fig:wall_layers}.  The other cases with $\sigma = -1$ and $\sigma = -3$  show ion concentrations that are indistinguishable after scaling the concentration with the case with $\sigma = -6$.   For ions near the wall there are two length scales associated with the ion layers.  The first length scale is the location of the closest ion layer to the wall, which occurs at the minimum of the Lennard-Jones potential of equation~\ref{eqn:phi_lj93}, at $\ell_* = \left(  18/45\right)^{1/6}b_w$ =  0.43 nm.  From the steric interactions the next closest layer can form only around $\ell_2 = \ell_* + b_{+} + b_{-}$.  For the parameters in Table~\ref{table:defaultParams} we have $\ell_2 = 0.66$ nm.  We see both of these length-scales manifest in the structure of the ion layers.  The double-layer essentially forms according to the packing distance imposed by the ion and wall sterics. This becomes especially pronounced as the concentration increases as seen in Figure ~\ref{fig:wall_layers}.    

Other interesting features arise in the ion layers near the wall as the ion concentrations increase.  The ion layers become smaller in width and more dense as the ion concentration increases.  For small concentrations there is significant overlap between the counterion and coion layers with significant mixing of ions especially within the secondary coion layer.  As the concentration increases the layers become more distinct.  Interestingly, for the counterion layer depletion occurs for the counterions within the secondary layer relative to the counterion concentration in the bulk.  This is especially pronounced once $C_m > 4$ as shown in the inset in Figure~\ref{fig:wall_layers}.  For $C_m < 4$ the concentration of the counterions appear to monotonically decay to the bulk counterion concentration.

\begin{figure}[H]
\centering
\includegraphics[width=0.99\columnwidth]{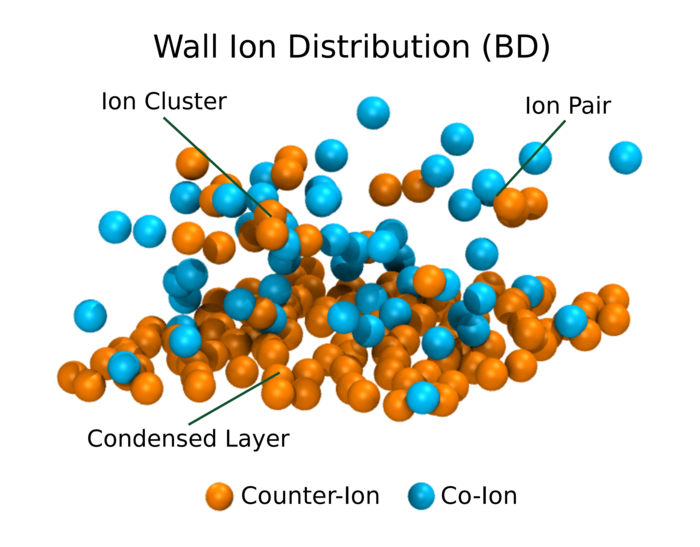}\caption{Ion configurations near the wall, for $\sigma = -6$ and $C_m = 8$.  }
\label{fig:wall_ion_3D}	
\end{figure}

In the nanochannel in the regimes we consider the ion double-layer structure is in contrast to many theories developed for weakly charged systems with a proposed stern layer and Helmholtz plane demarcating a transition from relatively immobile ions to a gaseous mobile phase of ions~\cite{KirbyBook2010,BazantBookChapter2011}.  From that perspective for our system at high ion concentrations this transition effectively occurs on the length scale of individual ions.  Near the wall the surface counterion and coion positions are strongly correlated, as shown in the simulation snapshot in Figure \ref{fig:wall_ion_3D}. Many of the ions form pairs with opposing ions or small clusters or chains.  The wall surface is covered in a condensed layer of counterions along with a secondary layer of coions that forms as part of clusters near individual counterions, see Figure~\ref{fig:wall_ion_3D}. This indicates some of the challenges involved in developing theory for such highly charged and concentrated regimes, where behaviors may be dependent on individual ion-ion interactions and charge clusters containing only a few ions.

\begin{figure}[H]
\centering
\includegraphics[width=0.99\columnwidth]{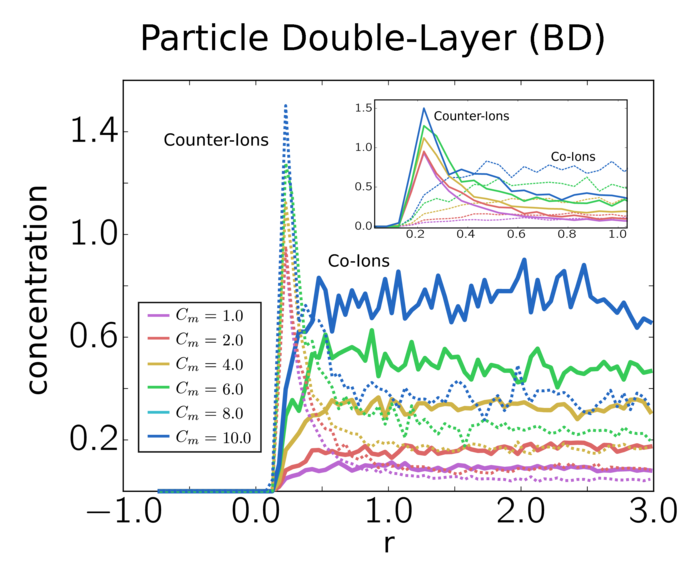}\caption{colloidal Particle Double-Layer  $(\sigma = -1.0)$. }
\label{fig:particle_layers_s_n1}	
\end{figure}

\begin{figure}[H]
\centering
\includegraphics[width=0.99\columnwidth]{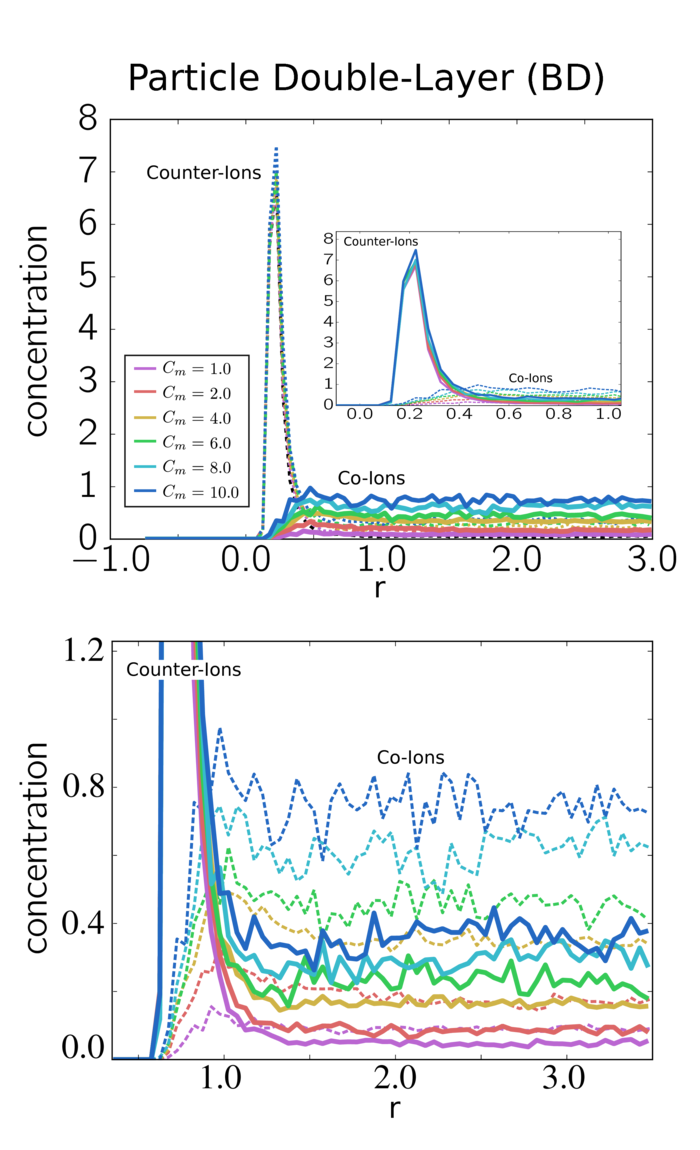}\caption{colloidal Particle Double-Layer  $(\sigma = -3.0)$.}
\label{fig:particle_layers_s_n3}	
\end{figure}

\begin{figure}[H]
\centering
\includegraphics[width=0.99\columnwidth]{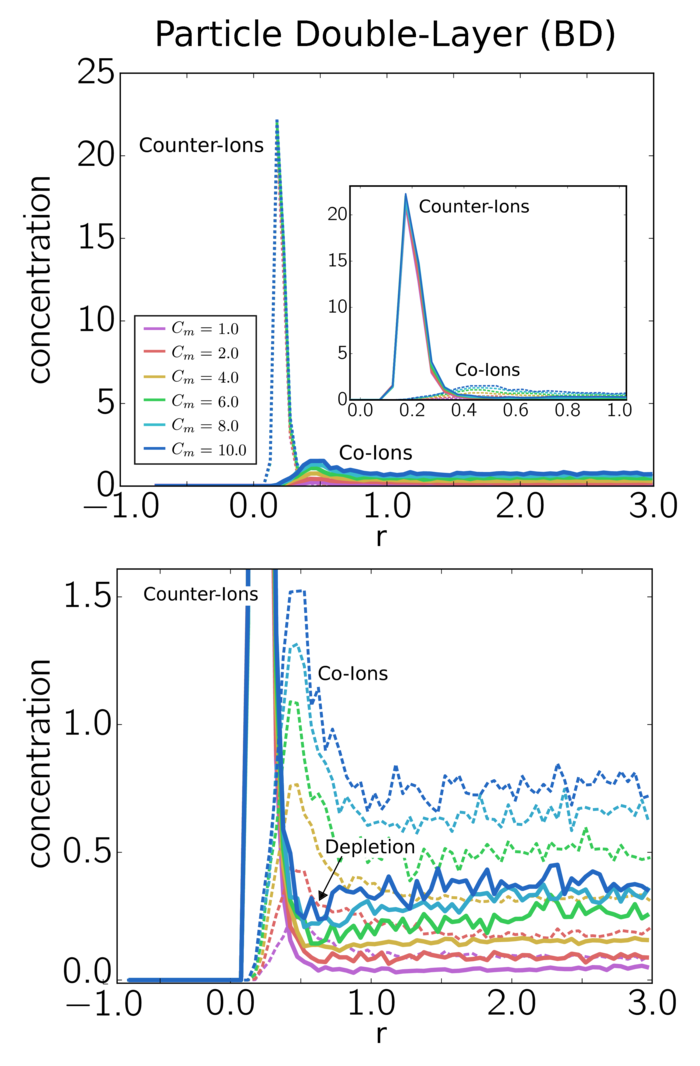}\caption{colloidal Particle Double-Layer $(\sigma = -6.0)$.}
\label{fig:particle_layers_s_n6}	
\end{figure}

Next we show the density of counterions and coions near the colloidal particle surface for the three different surface charges $\sigma = -1$, $\sigma = -3$, and $\sigma = -6$ in Figure~\ref{fig:particle_layers_s_n1}, Figure~\ref{fig:particle_layers_s_n3}, and Figure~\ref{fig:particle_layers_s_n6}.  The concentrations are measured at distances relative to the colloidal particle surface.  The relevant steric length-scale for the position of the counterion layer in this case is the steric length $\ell_{**} = 2^{1/6}(R+b) - R$ = 0.22 nm.  The coion layer forms at a distance corresponding to $\ell_{2*} = \ell_{**} + 2b_{\pm}= 0.45$ nm.  Again the layer locations are primarily determined by the packing of the ions as determined by the sterics.

For a relatively weak particle charge density of $\sigma = -1$, the counterions form a tight layer near the colloidal particle surface with significant mixing of coions into this primary layer.  After this layer the coions exhibit concentrations that rapidly approach a level comparable to the bulk, see Figure~\ref{fig:particle_layers_s_n1}. For $\sigma = -3$ the counterions also form a tight layer near the colloidal particle surface but with relatively little mixing of coions into this primary layer, see Figure~\ref{fig:particle_layers_s_n3}. The coions show only a weak secondary peak. For the highest surface charge density of $\sigma = -6$, a secondary layer of coions forms.  For the largest concentrations some depletion of the counterions is exhibited in the secondary layer relative to the bulk.  This is less pronounced than in the case of the walls due to the high curvature of the particle, but can be seen readily in the case with $\sigma = -6 e/nm^2$ and $C_m = 10$ as highlighted in the inset in Figure~\ref{fig:particle_layers_s_n6}.  

\begin{figure}[H]
\centering
\includegraphics[width=0.99\columnwidth]{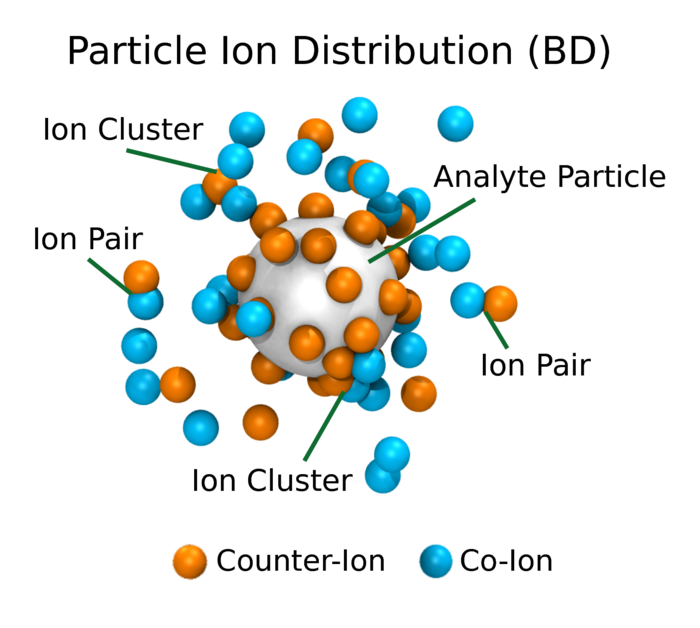}\caption{Ion configurations near the colloidal particle in bulk, for  $\sigma = -6$ and $C_m = 8$, showing ion pairs and clusters.}
\label{fig:particle_ion_3D}	
\end{figure}

\begin{figure}[H]
\centering
\includegraphics[width=0.99\columnwidth]{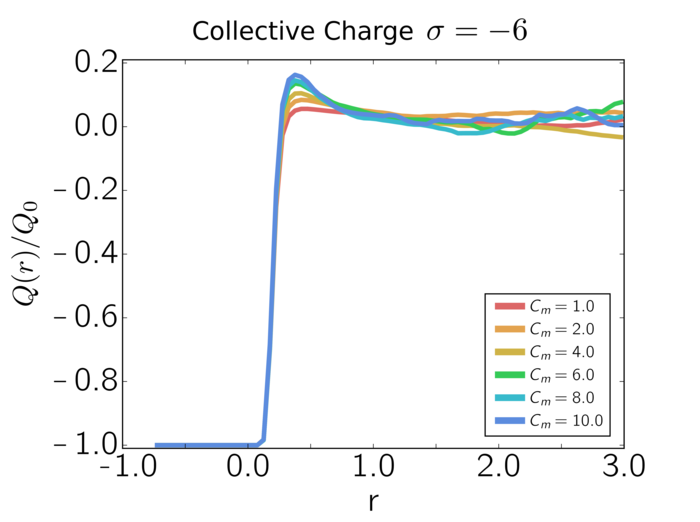}\caption{ The total collective amount of charge $Q(r)$ contained within the spherical volume of radius $r$ around the colloidal particle, for $\sigma = -6$.   Near the particle surface the double-layer provides excess charges (over-charging) when countering the colloidal particle charge. $Q_0$ is the colloidal particle charge.}
\label{fig:particle_overcharge}	
\end{figure}

For the smaller concentrations there is significant overlap of the counterion layer with the coion layer, with significant mixing in the secondary layer.  From examining configurations of the ions around the colloidal particle we find this arises from strong correlations between the counterions and coions resulting in the formation of transient charge clusters, as shown in Figure~\ref{fig:particle_ion_3D}.  As the colloidal particle charge increases, the layer of counterions near the particle adheres more strongly and the clusters are pushed increasingly toward the secondary layer.  For the case $\sigma = -6$ this is especially pronounced with the double-layer providing excess charge relative to what would be required to achieve local electric neutrality.  This over-charging phenomenon can be seen in Figure~\ref{fig:particle_overcharge}.

\subsection{Free Energy of colloidal Particle Location: BD Simulations}

We next consider the free energy $E(d)$ of the system as a function of  the colloidal particle position $d$, see Figure~\ref{fig:free_energy_all}.  The wall and the colloidal particle are both negatively charged, and  the free energy is  repulsive when the particle is sufficiently close to the wall.  As the concentration of the counterions and coions becomes sufficiently large, attraction occurs between the like-charged colloidal particle and wall.  The free energy minimum occurs at a distance comparable to the interaction length-scale of the first layers of ions of the wall and the colloidal particle surface.  The sum of the length-scale for the first counterion layer of the wall $\ell_* = 0.43nm$ and the length-scale of the counterion layer of the colloidal particle $\ell_{**} = 0.22nm$ is $\ell = \ell_* + \ell_{**} = 0.65 nm$, corresponding to $\bar{d}/\frac{1}{2}L \sim 0.22$, the approximate location of the free energy minima in Figure~\ref{fig:free_energy_all}. The free energy minimum can become significant compared to $k_B T$ at sufficiently large $C_m$. We discuss this further in Section~\ref{sec:discussion}.   
\multicolinterrupt{
\begin{figure}[H]
\centering
\includegraphics[width=1.0\columnwidth]{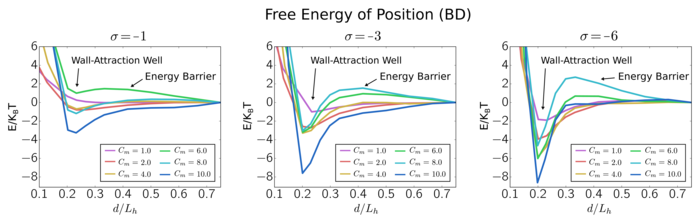}
\caption{Free energy profile of the colloid-wall distance from BD simulations.  We show the free energy for $\sigma$ = -1.0. -3.0, and -6.0, as a function of the distance $d$ between the particle and the wall.  The results are normalized by the thermal energy $k_B{T}$ and the half-width $L_h= \frac{1}{2}L$ of the nanochannel.  }
\label{fig:free_energy_all}	
\end{figure}
}

The free energy profile has an interesting non-monotonic dependence on the colloidal particle charge and electrolyte ion concentrations.  We see the depth of the free energy minimum well that forms near the wall is not entirely monotonic as the ionic concentration increases.  Most clearly, for $\sigma=-6$ the magnitude of the free energy well depth is larger for $C_m=6$ than for $C_m=8$, but then increases significantly for $C_m=10$.  There is also a significant free energy barrier  as large as $~2k_B{T}$ that can arise separating the particle from the free energy local minimum near the wall.  Making this even more interesting is that the largest energy barriers appear to occur for the intermediate ionic concentrations considered.  For instance see the cases with $\sigma = -3,-6$ and $C_m = 8$.  The free energy barrier appears to arise from the condensed ion layers that form on the colloidal particle surface and wall surface that must coordinate and rearrange as the particle approaches the wall, see Figure~\ref{fig:ions_particle_near_wall}.

\multicolinterrupt{

\begin{figure}[H]
\centering
\includegraphics[width=0.9\columnwidth]{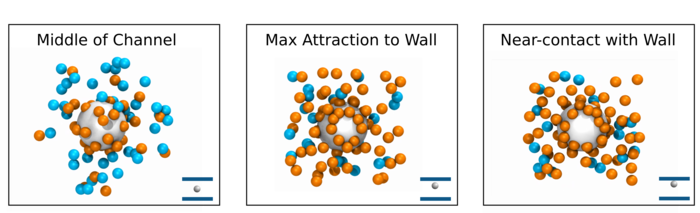}
\caption{Top-down view of colloidal particle and ion distribution, showing typical distributions of ions nearby the colloidal particle at different locations within the nanochannel.  We show the locations corrresponding to (i) the middle of the channel at $X_0^{(3)} = 3.0 nm$, (ii) the maximum attraction to the wall at $X_0^{(3)} = 4.6 nm$, and (iii) near-contact with the wall having large repulsion at $X_0^{(3)} = 4.85nm$.}
\label{fig:ions_particle_near_wall}	
\end{figure}
}

When the particle is at the free energy minimum, the counterions in the condensed layer typically form transient ring-like structures near the surface of the colloidal particle as shown in Figure~\ref{fig:ions_particle_near_wall}.  These counterions appear to serve double-duty in the condensed layer by screening both the colloidal particle charge and the effective wall charge.  This double-duty appears to be the source of the resulting free energy gain.  When the colloidal particle is positioned at an even closer distance to the wall it penetrates into the condensed counterion layer.  This excludes counterions which results in a significant pressure on the colloidal particle surface resulting in a strong free energy penalty.  It is important to remark that the effective electric field from the walls cancel so that all interactions beyond the steric distance are mediated by the ions.  

\subsection{Ion-Ion Correlations: BD Simulations}

To further understand the system, we examine the ion correlations in the condensed wall layer vs in the center of the channel. The counterions and coions exhibit strong self-correlations and cross-correlations.  The structures of these correlations depend significantly on whether an ion is near the channel wall or near the channel center.  As a matter of convention we refer to the ions near the channel center as being in the bulk.  We characterize the correlations by calculating a radial distribution function (RDF) $g(r)$ for ions within a permissible sampling region which we refer to as in the bulk or as near the wall (see Appendix~\ref{sec:corr_analysis} for details).  The RDFs $g(r)$ are normalized by the reference number concentration given by taking the count of all counterions or coions and dividing by the channel volume.  Throughout our simulations reference values are determined from the channel volume $V = 1944$ nm$^3$ and from the reference number concentrations $\hat{g}_- = 250/1944 \times C_m$ nm$^{-3}$ and $\hat{g}_+ = 150/1944 \times C_m$ nm$^{-3}$.  We remark that since the density of ions can be large near the walls the $g(r)$ can exhibit long-range normalized bulk values that are significantly less than $1.0$ and normalized wall values that are in excess of $1.0$.

The RDFs in the bulk are shown in Figure \ref{fig:ion_ion_correlations_bulk}. In the bulk, the counterion-counterion $g(r)$ shows a correlation hole, with the counterions not likely to be close together.  The counterion-coion interactions show strong correlations that indicate a counterion has a cluster of coions in its proximity at a distance roughtly twice the steric distance. The coion-coion $g(r)$'s exhibit a small  feature around $r = 0.5$ which appears to be related to ionic clusters that form with multiple coions associated to a common counterion. Since we have divalent counterions, it makes sense that there should roughly be two coions associated with each counterion. These results indicate that on average the bulk electrolyte consists of triples of ions with one counterion and two coions, but not larger ion clusters.

\multicolinterrupt{

\begin{figure}[H]
\centering
\includegraphics[width=0.99\columnwidth]{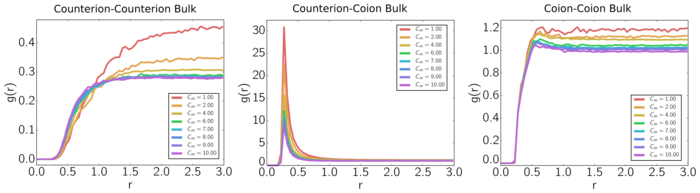}
\caption{Radial distribution functions $g(r)$ for ion-ion correlations in the bulk from the BD simulations.}
\label{fig:ion_ion_correlations_bulk}	
\end{figure}
}

Near to the wall, the RDF $g(r)$ exhibits features indicating much stronger correlations than in the bulk.  While the counterion-coion correlations are similar to those in the bulk, the counterion-counterion $g(r)$ has a significant peak at small $r$.  This is from the large density associated with the condensed counterion layer near the wall.  As the charge increases there is a transition around $C_m \geq 4$ from a correlated gas-like state to a state with significant correlations that are more liquid-like~\cite{bookChandler1987}.  The peak that develops moves closer toward the steric length-scale of the ions with peaks around $0.5$ nm.  The coion-coion correlations near the wall exhibit a peak for all of the regimes considered.  From examining simulation trajectories we find this arises from the strong correlations of the coions with the counterions and from bulk coions that transiently move to penetrate the strongly positively-charged condensed layer.  The coion-coion peak occurs independent of concentration around a similar length-scale $0.5$nm as the final counterion-counterion peaks for large concentration.  These results show that there are some significant differences in ion-ion correlations when near the wall relative to the bulk.

\multicolinterrupt{
\begin{figure}[H]
\centering
\includegraphics[width=0.99\columnwidth]{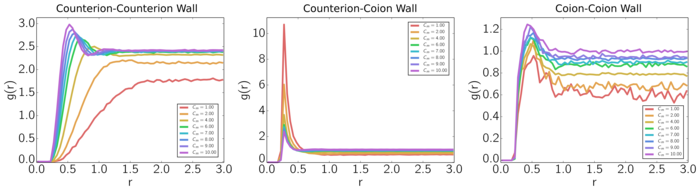}
\caption{Radial distribution functions $g(r)$ for ion-ion correlations near the wall from the BD simulations.}
\label{fig:ion_ion_correlations_wall}	
\end{figure}
}

\subsection{Results from Classical Density Functional Theory (cDFT) and Poisson-Boltzmann (PB) Theory}
The classical density functional theory (cDFT) and Poisson-Boltzmann (PB) theory provide other approaches for investigating phenomena in electrolytes and charged systems that are expected to be more computationally efficient than BD simulations.  However, in cDFT and PB further approximations are incurred in modeling the underlying physics of the charged system.  We expect that cDFT could provide a decent basis for describing the nanochannel system given the inclusion of terms accounting for charge correlations and ion sterics.  The steric and correlation effects can be seen in the ionic layering and clustered interactions in the simulation results particularly in Figures~\ref{fig:rpm_model} and~\ref{fig:particle_ion_3D}.  To further emphasize the importance of these effects, we include in our comparisons the mean-field Poisson-Boltzmann (PB) theory, which we do not expect to perform very well in the strongly charged regime.  These results further demonstrate the importance of ion correlation effects and sterics to obtain correct phenomenology even at a qualitative level.  As we shall discuss, our results further highlight the need for using descriptions beyond the mean-field theory to obtain reliable results in strongly charged regimes for the nanochannel system.

\begin{figure}[H]
\centering
\includegraphics[width=0.99\columnwidth]{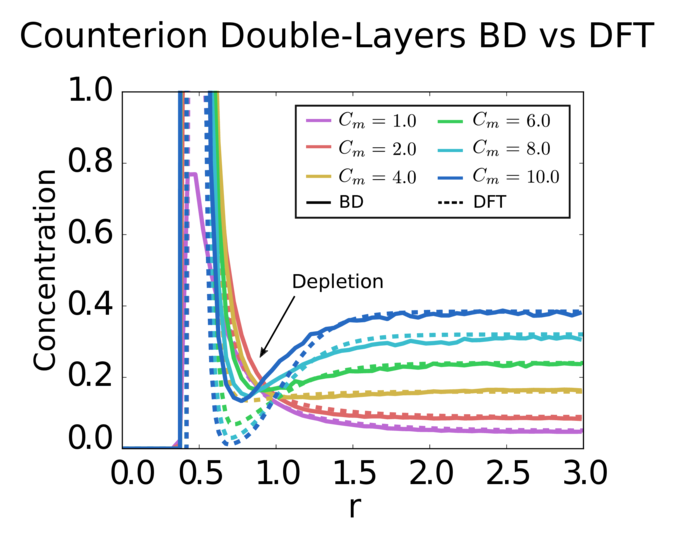}\caption{Comparison of the counterion densities for the cDFT (dashed curves) and the BD simulations (solid curves) as a function of distance $r$ from the channel wall, for wall charge densities from the $\sigma=-6$ column of Table \ref{table:wall_charge}.}
\label{fig:wallDL_DFT1}	
\end{figure}

\begin{figure}[H]
\centering
\includegraphics[width=0.99\columnwidth]{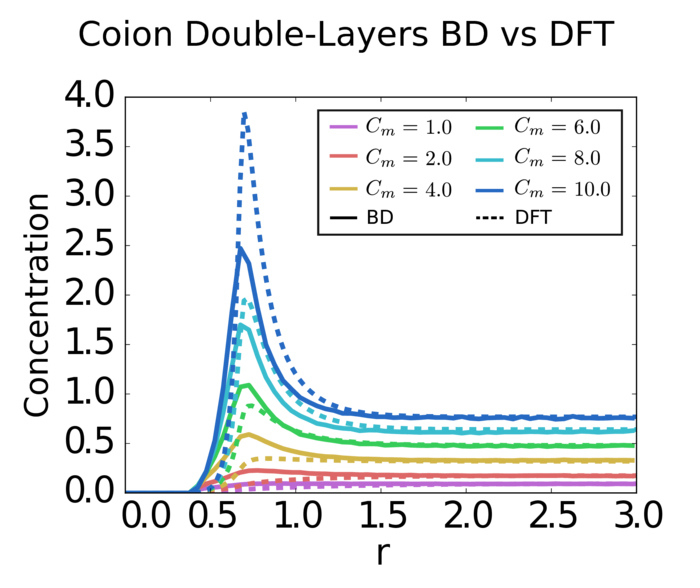}\caption{Comparison of the coion densities for the cDFT (dashed curves) and the BD simulations (solid curves) as a function of distance $r$ from the channel wall, for wall charge densities from the $\sigma=-6$ column of Table \ref{table:wall_charge}.}
\label{fig:wallDL_DFT2}	
\end{figure}

We compare the ion densities near the channel walls as calculated from cDFT with the simulation density profiles in Figures \ref{fig:wallDL_DFT1} and \ref{fig:wallDL_DFT2}.   We find that cDFT predicts qualitatively similar trends as  the simulations but with some significant quantitative differences.  At smaller values of $C_{m}$  the profiles exhibit monotonic behavior.  As observed in the BD simulation results, at larger values of $C_{m}$ the cDFT counterion densities exhibit a distinct peak (condensed layer) followed by a depleted region before attaining the bulk counterion concentration, see Figure~\ref{fig:wallDL_DFT1}.  The cDFT coion distributions exhibit a similar trend as in the BD results with a distinct peak occurring at the location of the depleted counterion region before attaining the bulk concentration, see Figure~\ref{fig:wallDL_DFT2}. The depletion after the first layer of counterions is not seen for ion densities calculated using the Poisson-Boltzmann equation, nor for cDFT calculations with only mean-field electrostatics (i.e., without the correlation term $F_{corr}$). Instead, in the absence of ion correlations, the counterions exhibit a single peak near the wall that decays monotonically to the bulk, whereas the coion density profiles simply increase monotonically from the wall to their bulk concentration, with no peak.

Thus, the cDFT charge correlation terms capture the charge density qualitatively as the ionic concentration is varied, but as the system becomes more strongly charged there are some significant quantitative deviations with the simulation results.  Compared to the BD simulations, at smaller $C_{m}$ the cDFT underestimates the magnitude of the coion peak but is in fairly good agreement with the long-range behavior of the counterion density profiles.  At larger values, $C_{m} > 6$, the cDFT overestimates the magnitude of the coion peak and also overestimates the amount of depletion in the counterion density. For all concentrations and wall charge densities, the cDFT overestimates the countertion contact density at the charged wall as compared with the BD simulations (not shown).

\begin{figure}[H]
\centering
\includegraphics[width=0.99\columnwidth]{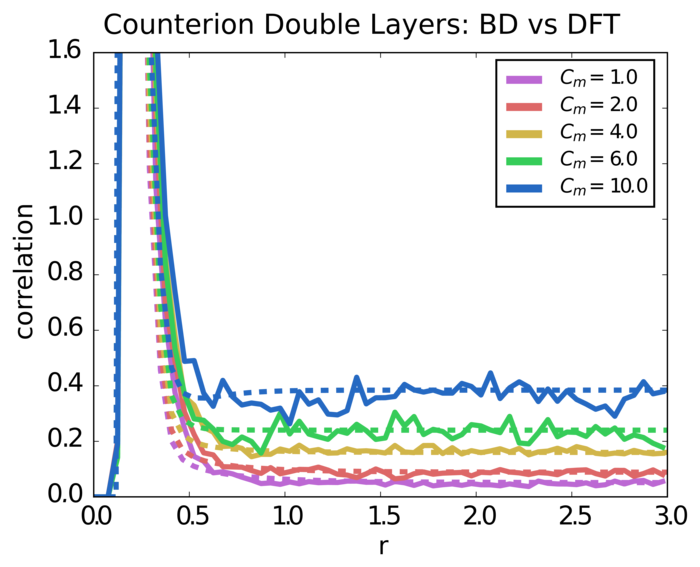}\caption{Comparison of the counterion densities for the cDFT (dashed curves) and the BD simulations (solid curves) as a function of distance $r$ from colloidal particle, for $\sigma=-3$.  }
\label{fig:macroDL_DFT1}	
\end{figure}

\begin{figure}[H]
\centering
\includegraphics[width=0.99\columnwidth]{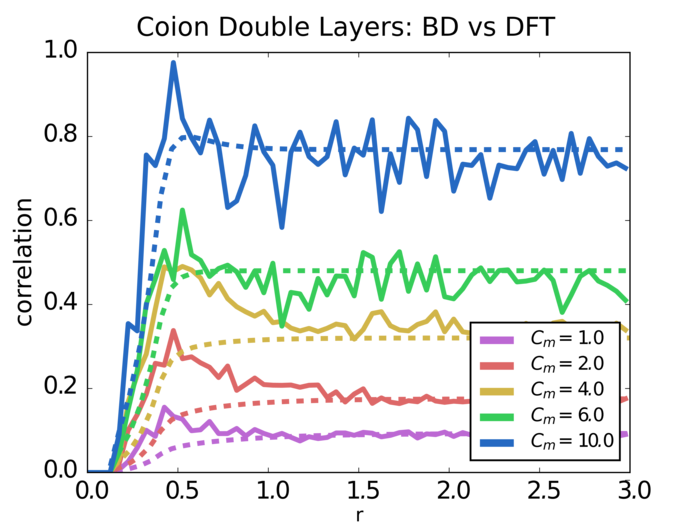}\caption{Comparison of the coion densities for the cDFT (dashed curves) and the BD simulations (solid curves) as a function of distance $r$ from the colloidal particle, for $\sigma=-3$. }
\label{fig:macroDL_DFT2}	
\end{figure}

Similar behavior is seen for the ion concentrations around the colloidal particle, as shown in Figures \ref{fig:macroDL_DFT1} and \ref{fig:macroDL_DFT2} for $\sigma = -3$. The cDFT underestimates the magnitude of the coion peak, especially for $C_{mult}=4$, and again overestimates the magnitude of the counterion contact density (not shown). 

We note that we are using the simplest form of the charge correlation term in the cDFT, namely the MSA expression for the direct correlation function $c(r)$, evaluated at the bulk density of the ions (i.e. the densities in the middle of the channel).  In our previous study of the interactions between charged nanoparticles in electrolyte, we found good agreement between cDFT and molecular dynamics simulations in the density profiles \cite{StevensFrischknechtNanoparticle2016}.  However, for our cDFT approach and for comparable regimes to our current studies,  discrepancies have been previously observed with simulations having large ion concentrations and in regions near to highly charged walls in the work of Oleksy and Hansen~\cite{Oleksy:2006ed}.  Oleksy and Hansen compared cDFT to Monte Carlo simulations for a 1:1 electrolyte at 1M concentration near a charged wall with reduced charge density $\sigma^* = 0.42$~\cite{Oleksy:2006ed}.  They also included a hard sphere solvent, and found differences in the ion density profiles of similar magnitude to those found in our work. Improvements to the charge correlation term, such as using the local weighted density in the calculation of $c(r)$, leads to excellent agreement between cDFT and e.g. molecular dynamics (MD) simulations near highly charged surfaces \cite{Lee:2012fj}.  The RFD functional of Gillespie and coworkers \cite{GillespieDensity2005}, which uses a local weighted density in $c(r)$, has been shown to give good agreement with simulation results and experiment in a variety of studies \cite{GillespieDensity2005,GillespiePennathurCorrelations2011}. Thus, in strongly charged regimes a more sophisticated approach beyond the simple bulk MSA treatment is needed to capture ion correlations if quantitative accuracy is sought near surfaces. In this paper, our main focus was to gain further insight into the qualitative role of charge correlations, so the more simple cDFT treatment is adequate. We also note that to our knowledge, more sophisticated treatments of charge correlations have not yet been implemented in a cDFT code that can also do 3D calculations in the geometry we study here.

\multicolinterrupt{
\begin{figure}[H]
\centering
\includegraphics[width=0.9\columnwidth]{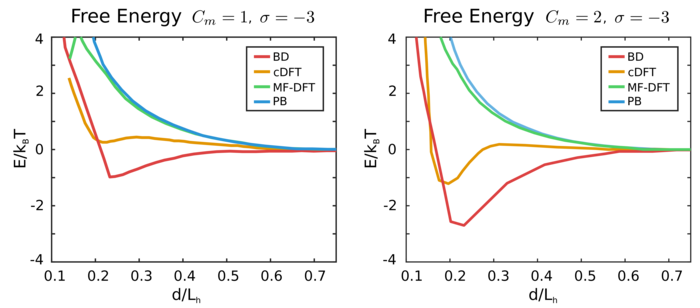}
\caption{Comparison of the free energy as a function of particle position in the nanochannel for cDFT, mean-field cDFT, and PB theory with the BD simulations.}
\label{fig:compare_rpm_cdft}	
\end{figure}
}

\begin{figure}[H]
\centering
\includegraphics[width=0.99\columnwidth]{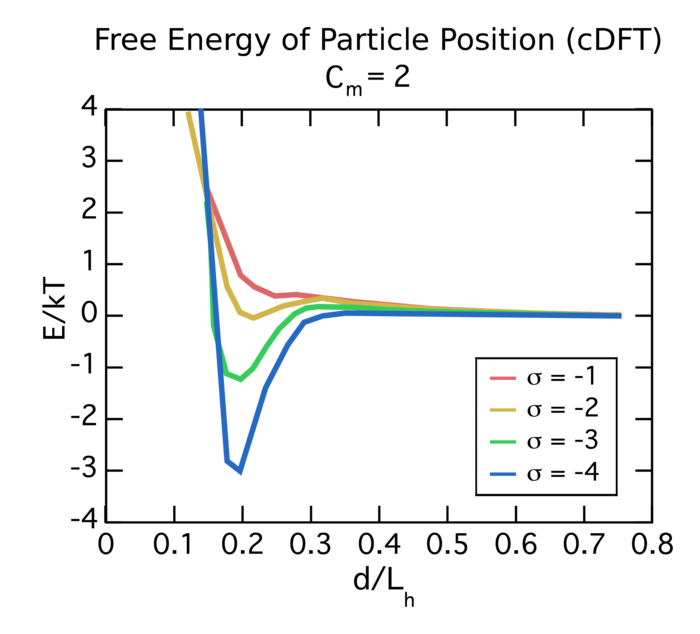}
\caption{Free energy of the colloidal particle as a function of position from cDFT as the particle charge $\sigma$ is varied.}
\label{fig:compare_cdft_free_eng}	
\end{figure}

Next we consider the free energy for the colloidal particle as a function of position in the nanochannel. For systems with large ionic concentrations and high charge density on the particle, the cDFT becomes computationally difficult to converge given the localized structures that develop within the density fields.  In Figure~\ref{fig:compare_rpm_cdft} we compare cDFT to the simulation results only for $\sigma = -3$ and $C_m = 1.0$ and $C_m = 2.0$, values which are  accessible with the cDFT computational methods.  We see that cDFT captures the trends on a qualitative level compared to the simulation results.  In particular, for sufficiently high charge,  the cDFT also predicts the development of a free energy minimum for the colloidal particle near the wall. In contrast, both the PB theory, which neglects sterics and correlations, and also mean-field cDFT with no charge correlations, are found to predict a purely repulsive interaction energy between the colloidal particle and wall. Figure~\ref{fig:compare_cdft_free_eng} shows cDFT results for differing charge densities on the colloidal particle, all at $C_m=2.0$. As the charge on the particle increases, the depth of the minimum in the free energy increases, as also found (for higher particle charges) by the BD simulations. In some cases the cDFT also predicts a small barrier in the free energy between the minimum and the center of the channel, but with cDFT we cannot access the high ion concentration regimes where this barrier is as large as in the BD simulations.

The difficulty in converging the cDFT calculations was surprising, but the systems studied here have higher ion concentrations and surface charge densities than most previous cDFT studies. In particular, our previous investigation of the interactions between like-charged nanoparticles had maximum ion concentrations of about 220 mM, which is close to the smallest ion concentration in the current study \cite{StevensFrischknechtNanoparticle2016}. Decreasing the strength of the electrostatic interactions slightly in the cDFT, by increasing the reduced temperature from $T^* = 0.33$ to $T^* = 0.43$, enabled convergence of systems with higher ion concentration (e.g., the $\sigma =-3$, $C_m=4$ system). This change corresponds to increasing the ion diameter from 0.232 nm to 3.0 nm. However, further increases would be needed in $T^*$ to get convergence at the higher ion concentrations so we did not pursue those calculations.

While cDFT agrees qualitatively with the simulation results there are some significant quantitative discrepancies.  The location of the free energy minimum in the cDFT is significantly closer to the nanochannel wall than in the BD simulations. This is likely due to the somewhat more narrow ion layers in the cDFT. We also find cDFT predicts a depth for the free energy well that is significantly smaller than observed in the simulation results, see Figure~\ref{fig:compare_rpm_cdft} and ~\ref{fig:compare_cdft_free_eng}.  Nevertheless, it is clear from these results that the attractive well results from ion charge correlations.

\section{Discussion}
\label{sec:discussion}

In the regimes studied, the ions tend to form clusters in the bulk electrolyte and a compact condensed layer near the channel walls.  The interplay between the ionic layers associated with the colloidal particle and the wall can result in a significant attraction between the like-charged colloidal particle and wall. As discussed in Section~\ref{sec:ion_dl_struct} this occurs at a distance comparable to the thickness of the condensed counterion layer.  As can be seen in Figure~\ref{fig:wall_layers} and~\ref{fig:particle_layers_s_n6}, there is a secondary layer of negative coions just beyond the counterion layer.  At the distance of the free energy minimum, the negatively charged colloidal particle joins the secondary layer of negative coions. From our comparisons between the BD simulations and the cDFT calculations, we found the attraction to be a consequence of the ion-ion correlations.  In contrast the mean-field theories, either PB or mean-field cDFT, that neglect these correlations predict a purely repulsive interaction between the colloidal particle and wall.  

The free energy of the colloidal particle location also exhibits an energy barrier.  For the case of the strongly charged colloidal particle and ion concentrations ($\sigma = -6$ and $C_m = 8$) there is a significant condensed counterion layer on the particle surface.  As the colloidal particle approaches the wall, the condensed layer of the colloidal particle merges with the condensed wall layer.  These rearrangements in some charge regimes result in the free energy barriers as observed in Figure~\ref{fig:free_energy_all}.  This effect appears to occur only for intermediate ion concentrations of $C_m = 6, 8$, for $\sigma=-3$ and -6, and disappears when the ion concentration becomes sufficiently large.  The significant rearrangements that occur as the ion approaches the wall indicate a strong role played by the ion-ion correlations and discrete structures in determining the free energy of the wall-particle interactions.

It is interesting to consider further the differences between multivalent and monovalent systems.  We performed two additional sets of simulations of monovalent systems with 1:1 electrolytes; further details and results are in Appendix \ref{sec:monovalent}. In the first set of simulations, we keep the number density of the monovalent ions the same as in the multivalent system.  While this case results in a different charge density it retains the same entropic contributions in the free energy.  In the second we keep the charge density of the system the same but double the number of counterions, which increases the number of charge carriers and the entropic contributions in the free energy.  In both cases, we find that the 1:1 electrolyte no longer results in a significant free energy minimum.  In the more strongly charged systems with more charge carriers the free energy minimum is further suppressed than in the case of the less charged system which shows a very small (relative to $k_B{T}$) and wide region of lower free energy, see Figure~\ref{fig:free_energy_all} and Figure~\ref{fig:mono_free_energy}.  This indicates that the multivalent system may benefit significantly from having fewer charge carriers, which reduces the entropic penalties associated with condensation of charge on the walls and strong correlations at the colloidal particle surface.  There is also more of an energy gain or less entropic loss when sharing a screening charge in common.  It can also be seen in the monovalent systems that the electrolyte is more diffuse, without the presence of transient ion clusters as in the multivalent system.  The simulation results indicate that it is the asymmetry between the ion charges and the reduced entropic penality for forming discrete structures that is responsible for the rich phenomena seen in multivalent electrolytes and charged systems.

Thus, the simulation results show that both the ion correlations, and the resulting discrete ion configurations, play important roles in determining the free energy of the system. In the BD simulations strong electrostatic interactions and multivalent ions can result in the formation of discrete clusters, and the interactions can be mediated at the level of individual ions and their arrangements, as seen in Figures~\ref{fig:rpm_model}, \ref{fig:particle_ion_3D} and \ref{fig:ions_particle_near_wall}.  This is expected to pose significant challenges in formulating constitutive equations for continuum descriptions of the system and in making quantitative predictions. The radial distribution functions we report for the counterions and coions for the bulk and near the wall may be helpful toward that aim, see Figure~\ref{fig:ion_ion_correlations_bulk} and~\ref{fig:ion_ion_correlations_wall}. The significant quantitative differences between the cDFT and the simulation results arise from the correlation terms in the cDFT functional that are based on the mean-spherical approximation (MSA) for bulk electrolytes. It would be of interest in future work to examine whether the RFD functional \cite{GillespieDensity2005}, which is still based on the MSA direct correlation function but for the local (inhomogeneous) rather than bulk density, would be sufficient to match the present simulation results, or whether improved expressions for the direct correlation function, such as from the new DH-extended MSA (DHEMSA) closure of Olvera de la Cruz and coworkers \cite{Zwanikken:2011jj}, would give better agreement. However, it may also be the case that for nano systems with finite numbers of ions, finite ion numbers lead to effects that cannot be captured by density functional theories which by construction only include the average ion density.

\section{Conclusion}
We have investigated the behaviors of a charged colloidal particles confined in nanochannels.  We have found for multivalent 2:1 electrolytes that strong ion-ion correlations can develop that give interesting free energy profiles for the colloidal particle position within the channel.  We found that the free energy profile can exhibit minima giving a preferred location for the colloid near the channel center and near to but separated from the channel wall.  We found in some of the charge regimes the minima can be separated by significant energy barriers.  This appears to be the result of over-charging of the double-layer that forms near the colloidal particle surface, see Figure~\ref{fig:particle_overcharge}.  Comparisons between our BD simulations and cDFT and PB theory indicate the strong role played by ion-ion correlations.  As may be expected from a mean-field, the PB theory was found to be inadequate in capturing even qualitative features of the simulation results.  The cDFT approach is found to capture at a qualitative level the main trends seen in the simulation results both for the ionic densities and for the free energy profile as the charge of the system is varied.  However, the cDFT results have quantitative discrepancies with the simulation results, in both the ionic layer densities near the walls and in the depth of the free energy well.  This arises appears to arise from the MSA approach used for the charge correlation term, which is based on hard-sphere models of unconfined bulk electrolytes.  Our simulations indicate that near surfaces the ions can form interesting ionic structures such as clusters or discrete layers differring significantly from bulk behaviors.   To obtain more quantitative accuracy, such effects would have to be captured likely requiring further development of correlation terms for cDFT.  Overall the cDFT did make predictions in qualitative agreement with most of the BD simulation results.

The results we report could have implications for many phenomena within nanochannels and more broadly nanodevices that rely upon electrical effects.  For instance, in the case of capillary electrophoresis the free energy profile indicates that colloidal particles within the device may hop between positions close to the nanochannel wall and close to the channel center.  Given the expected differences in particle mobilities in these locations, this could significantly affect arrival time observations.  More generally, our results show that discrete ion-ion interactions may play a dominate role in nanodevices requiring more sophisticated theory than proivided by traditional mean-field approaches such as the widely used Poisson-Boltzmann theory.  Toward this aim in developing better correlation terms for cDFT our bulk and wall radial distribution results may be useful.   Many of our results are expected to be useful in gaining insights into other charged systems such as biological macromolecules where similar discrete ion interactions and collective effects may be relevant.  

\section{Acknowledgments}
The authors P.J.A and I.S. acknowledge support from research grant NSF CAREER DMS-0956210, NSF DMS - 1616353, W. M. Keck Foundation, and DOE ASCR CM4 DE-SC0009254.  We also acknowledge UCSB Center for Scientific Computing NSF MRSEC (DMR-1121053) and UCSB MRL NSF CNS-0960316.  The authors would also like to thank Kai Sikorski for discussions and work developing codes for LAMMPS.  This work is supported by the Applied Mathematics Program within the Department of Energy (DOE) Office of Advanced Scientific Computing Research (ASCR) as part of the Collaboratory on Mathematics for Mesoscopic Modeling of Materials (CM4). This work was performed, in part, at the Center for Integrated Nanotechnologies, an Office of Science User Facility operated for the U.S. Department of Energy (DOE) Office of Science. Sandia National Laboratories is a multi-mission laboratory managed and operated by National Technology and Engineering Solutions of Sandia, LLC., a wholly owned subsidiary of Honeywell International, Inc., for the U.S. Department of Energy's National Nuclear Security Administration under contract DE-NA0003525.
\bibliographystyle{plain}
\bibliography{paperDatabase}{}

\begin{thebibliography}{10}

\bibitem{Allahyarov:1998kz}
E~Allahyarov, I~D'Amico, and H~Lowen.
\newblock {Attraction between Like-Charged Macroions by Coulomb Depletion}.
\newblock {\em Phys Rev Lett}, 81(6):1334--1337, 1998.

\bibitem{AtzbergerOsmosis2007}
Paul~J Atzberger and Peter~R Kramer.
\newblock Theoretical framework for microscopic osmotic phenomena.
\newblock {\em Phys Rev E Stat Nonlin Soft Matter Phys}, 75(6 Pt 1):061125, Jun
  2007.

\bibitem{BakerPNASElectrostatics2001}
Nathan~A. Baker, David Sept, Simpson Joseph, Michael~J. Holst, and J.~Andrew
  McCammon.
\newblock Electrostatics of nanosystems: {Application} to microtubules and the
  ribosome.
\newblock {\em Proceedings of the National Academy of Sciences},
  98(18):10037--10041, August 2001.

\bibitem{BaldessariDLOverlap2008}
Fabio Baldessari.
\newblock Electrokinetics in nanochannels: {Part} {I}. {Electric} double layer
  overlap and channel-to-well equilibrium.
\newblock {\em Journal of Colloid and Interface Science}, 325(2):526--538,
  September 2008.

\bibitem{Ballenegger:2009ct}
V~Ballenegger, A~Arnold, and J~J Cerd{\`a}.
\newblock Simulations of non-neutral slab systems with long-range electrostatic
  interactions in two-dimensional periodic boundary conditions.
\newblock {\em The Journal of Chemical Physics}, 131(9):094107--11, 2009.

\bibitem{BazantBookChapter2011}
Martin~Z. Bazant.
\newblock {\em Induced-Charge Electrokinetic Phenomena}, chapter~X, pages
  221--297.
\newblock Springer Vienna, Vienna, 2011.

\bibitem{BloomfieldDNACondensation1991}
Victor~A. Bloomfield.
\newblock Condensation of dna by multivalent cations: Considerations on
  mechanism.
\newblock {\em Biopolymers}, 31(13):1471--1481, 1991.

\bibitem{NetzInterfaces2016}
Douwe~Jan Bonthuis, Yuki Uematsu, and Roland~R. Netz.
\newblock Interfacial layer effects on surface capacitances and electro-osmosis
  in electrolytes.
\newblock {\em Philosophical Transactions of the Royal Society of London A:
  Mathematical, Physical and Engineering Sciences}, 374(2060), 2016.

\bibitem{bookChandler1987}
D.~Chandler.
\newblock {\em Introduction to Modern Statistical Mechanics}.
\newblock Oxford, New York, 1987.

\bibitem{DasDLOverlap2010}
Siddhartha Das and Suman Chakraborty.
\newblock Implications of {Interactions} between {Steric} {Effects} and
  {Electrical} {Double} {Layer} {Overlapping} {Phenomena} on
  {Electro}-{Chemical} {Transport} in {Narrow} {Fluidic} {Confinements}.
\newblock {\em arXiv:1010.5731 [cond-mat]}, October 2010.
\newblock arXiv: 1010.5731.

\bibitem{DerjaguinLandauColloids1941}
L.~Derjaguin, B.;~Landau.
\newblock Theory of the stability of strongly charged lyophobic sols and of the
  adhesion of strongly charged particles in solutions of electrolytes.
\newblock {\em Acta Physico Chemica URSS}, 633(14), 1941.

\bibitem{Evans:1979jn}
R~Evans.
\newblock {The nature of the liquid-vapour interface and other topics in the
  statistical mechanics of non-uniform, classical fluids}.
\newblock {\em Advances in Physics}, 28(2):143--200, April 1979.

\bibitem{FrischknechtCDFTNumerical2002}
L~J~D Frink, A~G Salinger, M~P Sears, J~D Weinhold, and A~L Frischknecht.
\newblock Numerical challenges in the application of density functional theory
  to biology and nanotechnology.
\newblock {\em Journal of Physics: Condensed Matter}, 14(46):12167, 2002.

\bibitem{Frink:2012hn}
Laura J~Douglas Frink, Amalie~L Frischknecht, Michael~A Heroux, Michael~L
  Parks, and Andrew~G Salinger.
\newblock Toward quantitative coarse-grained models of lipids with fluids
  density functional theory.
\newblock {\em J Chem Theory Comput}, 8(4):1393--1408, 2012.

\bibitem{Gardiner1985}
C.~W. Gardiner.
\newblock {\em Handbook of stochastic methods}.
\newblock Series in Synergetics. Springer, 1985.

\bibitem{Gelfand2000}
I.~M. Gelfand and S.~V. Fomin.
\newblock {\em Calculus of Variations}.
\newblock Dover, 2000.

\bibitem{GillespiePennathurCorrelations2011}
Dirk Gillespie, Aditya~S. Khair, Jaydeep~P. Bardhan, and Sumita Pennathur.
\newblock Efficiently accounting for ion correlations in electrokinetic
  nanofluidic devices using density functional theory.
\newblock {\em Journal of Colloid and Interface Science}, 359(2):520--529, July
  2011.

\bibitem{GillespieDensity2005}
Dirk Gillespie, Mónika Valiskó, and Dezso Boda.
\newblock Density functional theory of the electrical double layer: the {RFD}
  functional.
\newblock {\em Journal of Physics: Condensed Matter}, 17(42):6609, 2005.

\bibitem{GriffithsBookEM1998}
David~J. Griffiths.
\newblock {\em {Introduction to Electrodynamics (3rd Edition)}}.
\newblock Benjamin Cummings, 1998.

\bibitem{GronbechJensen:1998em}
Niels Gr{\o}nbech-Jensen, Keith~M Beardmore, and Philip Pincus.
\newblock {Interactions between charged spheres in divalent counterion
  solution}.
\newblock {\em Physica A}, 261(1-2):74--81, 1998.

\bibitem{RobbinsMultiscaleElectroOsmosis2016}
Lin Guo, Shiyi Chen, and Mark~O. Robbins.
\newblock Multi-scale simulation method for electroosmotic flows.
\newblock {\em The European Physical Journal Special Topics},
  225(8):1551--1582, 2016.

\bibitem{Hansen2000}
Jean-Pierre Hansen and Hartmut Löwen.
\newblock Effective interactions between electric double layers.
\newblock {\em Annual Review of Physical Chemistry}, 51(1):209--242, 2000.
\newblock PMID: 11031281.

\bibitem{Henderson:2011fn}
Douglas Henderson, Stanis{\l}aw Lamperski, Zhehui Jin, and J~Z Wu.
\newblock {Density Functional Study of the Electric Double Layer Formed by a
  High Density Electrolyte}.
\newblock {\em J Phys Chem B}, 115(44):12911--12914, 2011.

\bibitem{HerouxCDFTNumerical2007}
Michael~A. Heroux, Andrew~G. Salinger, and Laura J.~D. Frink.
\newblock Parallel segregated schur complement methods for fluid density
  functional theories.
\newblock {\em SIAM Journal on Scientific Computing}, 29(5):2059--2077, 2007.

\bibitem{Hockney1989}
Hockney and Eastwood.
\newblock {\em Computer Simulation Using Particles,}.
\newblock Adam Hilger, 1989.

\bibitem{PincusTwoPlates2009}
M~Kanduč, A~Naji, Y~S Jho, P~A Pincus, and R~Podgornik.
\newblock The role of multipoles in counterion-mediated interactions between
  charged surfaces: strong and weak coupling.
\newblock {\em Journal of Physics: Condensed Matter}, 21(42):424103, 2009.

\bibitem{KirbyBook2010}
Brian~J. Kirby.
\newblock {\em Micro-and nanoscale fluid mechanics: transport in microfluidic
  devices}.
\newblock Cambridge University Press, 2010.

\bibitem{KirbyZetaPotentialReview2004}
Brian~J. Kirby and Ernest~F. Hasselbrink.
\newblock Zeta potential of microfluidic substrates: 1. {Theory}, experimental
  techniques, and effects on separations.
\newblock {\em Electrophoresis}, 25(2):187--202, 2004.

\bibitem{SafinyaDNACondensation2000}
Ilya Koltover, Kathrin Wagner, and Cyrus~R. Safinya.
\newblock Dna condensation in two dimensions.
\newblock {\em Proceedings of the National Academy of Sciences},
  97(26):14046--14051, 2000.

\bibitem{Kuron2015}
Michael Kuron and Axel Arnold.
\newblock Role of geometrical shape in like-charge attraction of dna.
\newblock {\em The European Physical Journal E}, 38(3):20--, 2015.

\bibitem{GrierLikeChargeAttraction1997}
Amy~E. Larsen and David~G. Grier.
\newblock Like-charge attractions in metastable colloidal crystallites.
\newblock {\em Nature}, 385(6613):230--233, January 1997.

\bibitem{PincusChargeFluct2002}
A.~W.~C. Lau, D.~B. Lukatsky, P.~Pincus, and S.~A. Safran.
\newblock Charge fluctuations and counterion condensation.
\newblock {\em Phys. Rev. E}, 65(5):051502--, April 2002.

\bibitem{Lee:2012fj}
Jonathan~W. Lee, Robert~H Nilson, Jeremy~A. Templeton, Stewart~K Griffiths,
  Andy Kung, and Bryan~M. Wong.
\newblock {Comparison of Molecular Dynamics with Classical Density Functional
  and Poisson{\textendash}Boltzmann Theories of the Electric Double Layer in
  Nanochannels}.
\newblock {\em J Chem Theory Comput}, 8(6):2012--2022, 2012.

\bibitem{Maduar2016}
Salim~R. Maduar and Olga~I. Vinogradova.
\newblock Electrostatic interactions and electro-osmotic properties of
  semipermeable surfaces.
\newblock {\em The Journal of Chemical Physics}, 145(16):164703, 2016.

\bibitem{MANSOORI:1971p3814}
GA~Mansoori, NF~Carnahan, KE~Starling, and TW~Leland.
\newblock Equilibrium thermodynamic properties of misture of hard spheres.
\newblock {\em J Chem Phys}, 54(4):1523--{\&}, 1971.

\bibitem{McCammon1987}
J.~Andrew McCammon and Stephen~C. Harvey.
\newblock {\em Dynamics of Proteins and Nucleic Acids}.
\newblock Cambridge University Press, 1987.

\bibitem{NetzStrongCouplingTheory2000}
A.~G. Moreira and R.~R. Netz.
\newblock Strong-coupling theory for counter-ion distributions.
\newblock {\em EPL (Europhysics Letters)}, 52(6):705, 2000.

\bibitem{NetzSimilarChargedPlates2001}
Andre. Moreira and Roland~R. Netz.
\newblock Binding of similarly charged plates with counterions only.
\newblock {\em Phys. Rev. Lett.}, 87(7):078301--, July 2001.

\bibitem{BAM}
Richard~P. Muller, Randall~T. Cygan, Jie Deng, Amalie~L. Frischknecht, John~C.
  Hewson, Michael~P. Kanouff, Richard Larson, Harry~K. Moffat, Craig~M. Tenney,
  Peter~A. Schultz, and Gregory~J. Wagner.
\newblock Modeling thermal abuse in transportation batteries.
\newblock Technical Report SAND2012-7816, Sandia National Laboratories,
  Albuquerque, New Mexico and Livermore, California, 2012.

\bibitem{NagornyakLikeChargeExp2009}
Ekaterina Nagornyak, Hyok Yoo, and Gerald~H. Pollack.
\newblock Mechanism of attraction between like-charged particles in aqueous
  solution.
\newblock {\em Soft Matter}, 5:3850--3857, 2009.

\bibitem{Netz2000}
R.R. Netz and H.~Orland.
\newblock Beyond poisson-boltzmann: Fluctuation effects and correlation
  functions.
\newblock {\em The European Physical Journal E}, 1(2):203--214, Feb 2000.

\bibitem{Oleksy:2006ed}
Anna Oleksy and Jean-Pierre Hansen.
\newblock {Towards a microscopic theory of wetting by ionic solutions. I.
  Surface properties of the semi-primitive model}.
\newblock {\em Mol Phys}, 104(18):2871--2883, 2006.

\bibitem{bookReviewCompChemistry2015}
Abby~L. Parrill and Kenny Lipkowitz.
\newblock {\em Reviews in Computational Chemistry, Volume 28, page 226}.
\newblock Wiley, 2015.

\bibitem{PegadoLikeChargeAttraction2008}
Luís Pegado, Bo~Jönsson, and Håkan Wennerström.
\newblock Like-charge attraction in a slit system: pressure components for the
  primitive model and molecular solvent simulations.
\newblock {\em Journal of Physics: Condensed Matter}, 20(49):494235, 2008.

\bibitem{PennathurTransport2004}
Sumita Pennathur and Juan~G. Santiago.
\newblock Transport {Mechanisms} in {Electrokinetic} {Nanoscale} {Channels}.
\newblock {\em Fluids Engineering Conference}, pages 191--196, January 2004.

\bibitem{PlimptonLAMMPS1995}
Steve Plimpton.
\newblock Fast parallel algorithms for short-range molecular dynamics.
\newblock {\em Journal of Computational Physics}, 117(1):1 -- 19, 1995.

\bibitem{PollockPPPM1996}
E.L. Pollock and Jim Glosli.
\newblock Comments on p3m, fmm, and the ewald method for large periodic
  coulombic systems.
\newblock {\em Computer Physics Communications}, 95(2):93 -- 110, 1996.

\bibitem{bookHandbookEng2007}
Andrei~D. Polyanin and Alexander~V. Manzhirov.
\newblock {\em Handbook of Mathematics for Engineers and Scientists}.
\newblock Taylor \& Francis, 2007.

\bibitem{Roth:2002p518}
R~Roth, R~Evans, A~Lang, and G~Kahl.
\newblock Fundamental measure theory for hard-sphere mixtures revisited: the
  white bear version.
\newblock {\em J Phys-Condens Mat}, 14(46):12063--12078, 2002.

\bibitem{Ryzsko2010}
W.~Ryzsko, A.~Patrykiejew, S.~SokoÅowski, and O.~Pizio.
\newblock Phase behavior of a two-dimensional and confined in slitlike pores
  square-shoulder, square-well fluid.
\newblock {\em The Journal of Chemical Physics}, 132(16):164702, 2010.

\bibitem{SaderAttractionUnresolved1999}
John~E. Sader and Derek~Y.C. Chan.
\newblock Long-range electrostatic attractions between identically charged
  particles in confined geometries: An unresolved problem.
\newblock {\em Journal of Colloid and Interface Science}, 213(1):268 -- 269,
  1999.

\bibitem{StevensFrischknechtNanoparticle2016}
K.~Michael Salerno, Amalie~L. Frischknecht, and Mark~J. Stevens.
\newblock Charged nanoparticle attraction in multivalent salt solution: A
  classical-fluids density functional theory and molecular dynamics study.
\newblock {\em J. Phys. Chem. B}, 120(26):5927--5937, July 2016.

\bibitem{PincusCondensationPolyelectrolytes1998}
H.~Schiessel and P.~Pincus.
\newblock Counterion-condensation-induced collapse of highly charged
  polyelectrolytes.
\newblock {\em Macromolecules}, 31(22):7953--7959, 1998.

\bibitem{SquiresQuakeFluidicsReview2005}
Todd~M. Squires and Stephen~R. Quake.
\newblock Microfluidics: Fluid physics at the nanoliter scale.
\newblock {\em Rev. Mod. Phys.}, 77(3):977--, October 2005.

\bibitem{StevensDNACond2001}
M~J Stevens.
\newblock Simple simulations of dna condensation.
\newblock {\em Biophysical Journal}, 80(1):130--139, January 2001.

\bibitem{McCammonBoLiSCPF2015}
Hui Sun, Jiayi Wen, Yanxiang Zhao, Bo~Li, and J.~Andrew McCammon.
\newblock A self-consistent phase-field approach to implicit solvation of
  charged molecules with poissonâboltzmann electrostatics.
\newblock {\em The Journal of Chemical Physics}, 143(24):243110, 2015.

\bibitem{TorrieRPM_MC_1979}
G.M. Torrie and J.P. Valleau.
\newblock A monte carlo study of an electrical double layer.
\newblock {\em Chemical Physics Letters}, 65(2):343 -- 346, 1979.

\bibitem{ValleauRPM_ElectrolytesI1980}
John~P. Valleau and L.~Kenneth Cohen.
\newblock Primitive model electrolytes. i. grand canonical monte carlo
  computations.
\newblock {\em The Journal of Chemical Physics}, 72(11):5935--5941, 1980.

\bibitem{ValleauRPMElectrolytesII1980}
John~P. Valleau, L.~Kenneth Cohen, and Damon~N. Card.
\newblock Primitive model electrolytes. ii. the symmetrical electrolyte.
\newblock {\em The Journal of Chemical Physics}, 72(11):5942--5954, 1980.

\bibitem{OverbeekColloids1948}
J.~Th.~G. Verwey, E. J. W.;~Overbeek.
\newblock {\em Theory of the stability of lyophobic colloids}.
\newblock Amsterdam: Elsevier., 1948.

\bibitem{wanDLOverlap1997}
Qian-Hong Wan.
\newblock Effect of {Electrical} {Double}-{Layer} {Overlap} on the
  {Electroosmotic} {Flow} in {Packed}-{Capillary} {Columns}.
\newblock {\em Analytical Chemistry}, 69(3):361--363, February 1997.

\bibitem{AtzbergerLAMMPS2016}
Y.~Wang, J.~Sigurdsson, and P.~Atzberger.
\newblock Fluctuating hydrodynamics methods for dynamic coarse-grained
  implicit-solvent simulations in lammps.
\newblock {\em SIAM J. Sci. Comput.}, 38(5):S62--S77, January 2016.

\bibitem{WeeksChandlerAndersen1971}
John~D. Weeks, David Chandler, and Hans~C. Andersen.
\newblock Role of repulsive forces in determining the equilibrium structure of
  simple liquids.
\newblock {\em J. Chem. Phys.}, 54(12):5237--5247, June 1971.

\bibitem{AtzbergerWuPeskinOsmosisVesicle2015}
Chen-Hung Wu, Thomas~G. Fai, Paul~J. Atzberger, and Charles~S. Peskin.
\newblock Simulation of osmotic swelling by the stochastic immersed boundary
  method.
\newblock {\em SIAM Journal on Scientific Computing}, 37(4):B660--B688, 2015.

\bibitem{XingWallsPB2011}
Xiangjun Xing.
\newblock Poisson-{Boltzmann} theory for two parallel uniformly charged plates.
\newblock {\em Physical Review E}, 83(4):041410, April 2011.

\bibitem{Yeh:1999dm}
In-Chul Yeh and Max~L Berkowitz.
\newblock {Ewald summation for systems with slab geometry}.
\newblock {\em Journal of Chemical Physics}, 111(7):3155--3162, 1999.

\bibitem{Yu:2002kx}
Yang-Xin Yu and J~Z Wu.
\newblock {Structures of hard-sphere fluids from a modified fundamental-measure
  theory}.
\newblock {\em J Chem Phys}, 117(22):10156--10164, 2002.

\bibitem{Zwanikken:2011jj}
Jos~W Zwanikken, Prateek~K Jha, and Monica~Olvera de~la Cruz.
\newblock {A practical integral equation for the structure and thermodynamics
  of hard sphere Coulomb fluids}.
\newblock {\em J Chem Phys}, 135(6):064106, 2011.

\end{thebibliography}

\clearpage
\newpage

\appendix

\section{Classical Density Functional Theory (cDFT) Formulation} 
\label{sec:detailsDFT}

We provide here some additional discussion and details concerning our formulation of the cDFT.  As we discussed in Section~\ref{sec:cDFT}, the Helmholtz free energy 
consists of the terms:
\begin{eqnarray}
F\left[  \rho_{\alpha}(\mathbf{r}) \right]  
& = &
F_{id}\left[  \rho_{\alpha}(\mathbf{r}) \right] 
+ F_{hs}\left[  \rho_{\alpha}(\mathbf{r})  \right] \\ 
\nonumber
& + &   F_{coul}\left[  \rho_{\alpha}(\mathbf{r}) \right] 
+ F_{corr}\left[  \rho_{\alpha}(\mathbf{r}) \right ].
\label{eq:helm}
\end{eqnarray}
The terms represent respectively the Helmholtz free energies for the ideal gas (id), hard spheres (hs), mean-field Coulombic interactions (coul), and second order charge correlations (corr).  The term $F_{id}$ is the free energy of an ideal gas which incorporates the translational free energy as
\begin{eqnarray}
F_{id}[\rho_\alpha({\bf r})] \hspace{6cm} \\ 
\nonumber
= k_BT \sum_\alpha \int d{\bf r}  
\rho_\alpha({\bf r}) \left[ \ln (\Lambda_\alpha^3 \rho_\alpha({\bf r})) -1\right].
\end{eqnarray}
Here the thermal de Broglie wavelengths $\Lambda_\alpha$ are constants throughout and do not influence the free energy of the system, so they will be neglected.

For the hard sphere contribution $F_{hs}$ we use the fundamental measure theory of \cite{Roth:2002p518,Yu:2002kx} given by
\begin{equation} 
\label{equ:cdft_hs}
F_{hs}\left[  \rho_{\alpha}(\mathbf{r}) \right]=  k_BT \int{d\mathbf{r} \,
                       \Phi[n_{\gamma}(\mathbf{r})]}.
\end{equation}
The energy density for the hard sphere system $\Phi$ is a functional of the Rosenfeld nonlocal (weighted) densities $n_{\gamma}$ given by
\begin{eqnarray}
\label{eq:PhiWB}
\Phi & = & -n_{0}\ln\left(1-n_{3} \right) \hspace{2cm} \\
\nonumber
& + & \frac{n_{1}n_{2} - n_{V1}\cdot n_{V2}}{1-n_{3}} \hspace{1.23cm} \\
\nonumber
& + & \left(n_2^3 - 3n_2 n_{V2}\cdot n_{V2}\right)\cdot \\ 
\nonumber                        
& \cdot &                          \frac{n_3+(1-n_3)^2\ln(1-n_3)}{36\pi n_{3}^{2}(1-n_{3})^2}. 
\end{eqnarray}  
The nonlocal densities are
\begin{equation}                       
n_{\gamma}(\mathbf{r})=\sum_{\alpha}\int d\mathbf{r}' \,  \rho_{\alpha}(\mathbf{r})
                \omega_{\alpha}^{(\gamma)}(\mathbf{r} - \mathbf{r}'),
\end{equation}
where $\omega_{\alpha}^{(\gamma)}$ are the weight functions.  The weight functions are based on geometric properties of the interactions between hard spheres and are given by the specific forms
\begin{align}
\notag
       \omega_{\alpha}^{(2)}(\mathbf{r}) &= \delta(R_{\alpha}-|\mathbf{r}|), 
           & \omega_{\alpha}^{(3)}(\mathbf{r}) &= \theta(R_{\alpha}-|\mathbf{r}|), \\ \notag
       \omega_{\alpha}^{(0)}(\mathbf{r}) &= \frac{ \omega_{\alpha}^{(2)}(\mathbf{r}) }{4\pi R_{\alpha}^2},
           & \omega_{\alpha}^{(1)}(\mathbf{r}) &=  \frac{ \omega_{\alpha}^{(2)}(\mathbf{r}) }{4\pi R_{\alpha}}, \\ \notag
       \omega_{\alpha}^{(V2)}(\mathbf{r}) &= \frac{\mathbf{r}}{r}\delta(R_{\alpha}-|\mathbf{r}|), 
           & \omega_{\alpha}^{(V1)}(\mathbf{r})
               &= \frac{\omega_{\alpha}^{(V2)}(\mathbf{r})}{4\pi R_{\alpha}}. \notag \\
               \label{eqn:cdft_hs_weights}
\end{align}
The $\delta(\mathbf{r})$ denotes the Dirac delta function and the $\theta(\mathbf{r})$ denotes the Heaviside step function. The functional consisting of equation~\ref{equ:cdft_hs} - \ref{eqn:cdft_hs_weights} is designed to match the Mansoori-Carnahan-Starling-Leland (MCSL) equation of state for multi-component hard-sphere fluids~\cite{MANSOORI:1971p3814}.

The contribution to the free energy $F_{coul}$ accounts for
the mean-field part of the electrostatic interactions as
\begin{eqnarray}
\\
\nonumber
F_{coul}\left[  \rho_{\alpha}(\mathbf{r}) \right]  \hspace{5.5cm}  \\
\nonumber
= \frac{1}{2} \sum_{\alpha\beta} \int d{\bf r} \int d{\bf r'} \rho_\alpha({\bf r}) \rho_\beta({\bf r'})\frac{q_\alpha q_\beta}{4\pi\epsilon_0 \epsilon |{\bf r-r'}|} \\
 \nonumber
	 =  \frac{1}{2} \sum_\alpha \int d{\bf r} q_\alpha \rho_\alpha({\bf r}) \phi({\bf r}).  \hspace{3cm}
\end{eqnarray}
Here $q_\alpha$ is the charge of species $\alpha$, $\epsilon_0$ is the permittivity of free space, and $\epsilon$ denotes the relative dielectric constant.  We introduce the electrostatic potential $\phi({\bf r})$ in the second expression.

The contribution to the free energy $F_{corr}$ accounts for the charge correlations of the electrostatic interactions.  We use for the charge correlation the approach in~\cite{Oleksy:2006ed} with
\begin{eqnarray}
	\label{eq:deltac}
\\
\nonumber
F_{corr}\left[  \rho_{\alpha}(\mathbf{r}) \right] \hspace{5.0cm} \\
\nonumber
= -\frac{1}{2} k_BT \sum_{\alpha\beta} \int d{\bf r} \int d{\bf r'} \hspace{2.5cm} \\
\nonumber
\rho_\alpha({\bf r}) \rho_\beta({\bf r'}) \Delta c_{\alpha\beta}(|{\bf r-r'}|).
\end{eqnarray}
The correlation operator is
\begin{eqnarray}
	\label{eq:deltac2}
\Delta c_{\alpha\beta}(|{\bf r-r'}|) 
& = & c_{\alpha\beta}(r) \\
\nonumber
& + & \frac{q_\alpha q_\beta}{4\pi \epsilon_0\epsilon k_BT|{\bf r-r'}|}  \\
\nonumber
& - & c_{\alpha\beta}^{HS}(r).
\end{eqnarray}
where $c_{\alpha\beta}(r)$ is the direct correlation function for the bulk charged system~\cite{Oleksy:2006ed}.  The hard sphere and Coulombic terms are subtracted from the full direct correlation function $c_{\alpha\beta}(r)$ in equation~\ref{eq:deltac2} to avoid double counting relative to the contributions already in the $F_{hs}$ and $F_{coul}$ terms. The form of $c_{\alpha\beta}(r)$ is taken from the known analytic solution of the mean-spherical-approximation (MSA) for a mixture of charged hard spheres.  Detailed expressions can be found in the reference~\cite{Oleksy:2006ed}.

The grand free energy for the density field of equation~\ref{eq:omega1} is minimized by solving an associated set of Euler-Lagrange equations.  This is formulated in terms of residuals with the objective of obtaining densities so that $R_i = 0$.  The numerical methods used and other computational details can be found in discussion of the Tramonto package in~\cite{FrischknechtCDFTNumerical2002,
HerouxCDFTNumerical2007,BAM,Frink:2012hn}.  The residuals are given by
\begin{eqnarray}
	R_1 & = & \ln \rho_\alpha({\bf r}) + V({\bf r}) - \mu_\alpha \\
		\nonumber
	& + & \int \sum_\gamma \frac{\partial \Phi}{\partial n_\gamma} ({\bf r'}) \omega_{\alpha}^{(\gamma)}(\mathbf{r} - \mathbf{r}') d {\bf r'}  \\
	\nonumber
	& + & \sum_\beta \int d{\bf r'} \rho_\beta({\bf r'}) u_{\alpha\beta}({\bf r-r'}) \\
		\nonumber
	& - & \sum_\beta \int d{\bf r'} \rho_\beta({\bf r'}) \Delta c_{\alpha\beta}({\bf r-r'}) + Z_\alpha \phi({\bf r})
	\label{eq:r1}
\end{eqnarray}
\begin{eqnarray}
\\
\nonumber
R_2 & = & n_\gamma({\bf r}) - \sum_{\alpha}\int d\mathbf{r}' \,  \rho_{\alpha}(\mathbf{r})
\omega_{\alpha}^{(\gamma)}(\mathbf{r} - \mathbf{r}') \\
R_3 & = & \nabla^2 \phi -\frac{4\pi}{T^*} 
\sum_\alpha q_\alpha \rho_\alpha.
		\label{eq:r3}
\end{eqnarray}
In these expressions we have adopted the convention that all quantities are in reduced units, so energies are in units of $k_BT$, lengths in units of $d$, and valence in terms of $Z_\alpha$ for species $\alpha$.  Additional information concerning classical Density Functional Theory (cDFT) in general can be found in~\cite{Oleksy:2006ed,Henderson:2011fn} and our specific approach to cDFT in~\cite{FrischknechtCDFTNumerical2002,
HerouxCDFTNumerical2007,BAM,Frink:2012hn}.

\section{Monovalent Ion Correlations}
\label{sec:monovalent}
\begin{figure}[H]
\centering
\includegraphics[width=0.99\columnwidth]{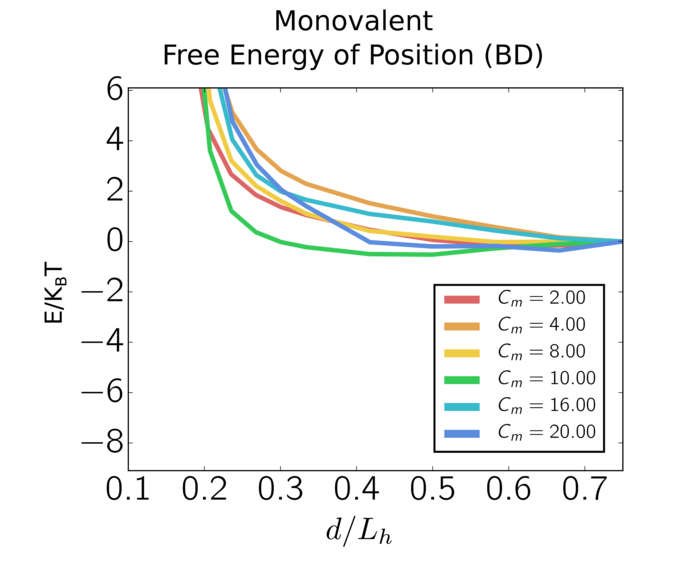}
\caption{Free Energy colloidal Particle Position.   Monovalent case with $\sigma = -6$, $C_m = 2,4,8,10,16,20$.}
\label{fig:mono_free_energy}	
\end{figure}

We performed additional BD simulations for the nanochannel system with a monovalent 1:1 electrolyte with the conditions that $\sigma = -6$ and $C_m = 2,8,10$ and $C_m = 4,16,20$.  This allows us to make comparisons with the multivalent cases when changing either the total charge of the system or while keeping charge fixed and changing only the number of charge carriers for the counterions.  We report the free energy for the colloidal particle position for constant number density in Figure~\ref{fig:mono_free_energy}.  We report the ion-ion correlations and radial distribution function $g(r)$ for ions in the bulk and near the wall in Figure~\ref{fig:mono_ion_ion_correlations_bulk} and~\ref{fig:mono_ion_ion_correlations_wall}.  We discuss the $g(r)$ analysis to distinguish these regions in Appendix~\ref{sec:corr_analysis}.

We find for all of the monovalent cases that there is no significant free energy minimum that forms for a preferred location for the colloidal particle within the channel, see Figure~\ref{fig:mono_free_energy}.  This is in contrast to the free energy minima in comparable regimes seen in Figure~\ref{fig:free_energy_all}.  It is interesting to note that the case with $C_m = 10$ shows some free energy reduction as the colloidal particle approaches the wall but it is insignificant relative to $k_B{T}$.  From observations of the simulation trajectory one can see again significant ion condensation on both the walls and the colloidal particle surface.  A mechanism similar to that discussed in Section~\ref{sec:discussion} may be at play but it appears the free energy gain is much reduced by the strength of the individual ion charges and entropic penalty associated with monovalent ions.

We further explore the ion-ion correlations in the monovalent cases.  We find that there are correlations between the individual counterions and coions as one may expect.  However, in the bulk there is little to no coordination in the counterion-counterion or coion-coion interactions, see Figure~\ref{fig:mono_ion_ion_correlations_bulk}.  Near the walls, while we find there is little to no coordination in the counterion-counterion interactions there is some significant coordination in the coion-coion interactions, see Figure~\ref{fig:mono_ion_ion_correlations_wall}.   From examination of the simulation trajectory of the system this appears to arise from the transient insertion of coions into the counterion-rich condensed layer near the walls.  In contrast to the multivalent case we find for the monovalent electrolyte there are not significant ion clusters or other discrete ion structures that form in the bulk electrolyte.  

Finally, cDFT calculations for monovalent electrolyte with $\sigma=-3$ and $C_m = 2, 4$ also show a monotonically increasing free energy as the colloidal particle nears the channel wall, in agreement with the simulations.

\multicolinterrupt{

\begin{figure}[H]
\centering
\includegraphics[width=0.99\columnwidth]{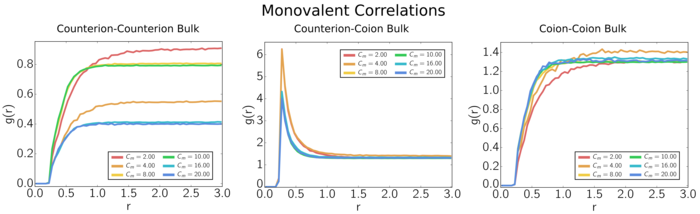}
\caption{Ion Correlations in the Bulk.  The RDF $g(r)$ for ion-ion correlations in proximity to the wall for the  monovalent case with $\sigma = -6$, and $C_m = 2,4,8,10,16,20$.}
\label{fig:mono_ion_ion_correlations_bulk}	
\end{figure}

\begin{figure}[H]
\centering
\includegraphics[width=0.99\columnwidth]{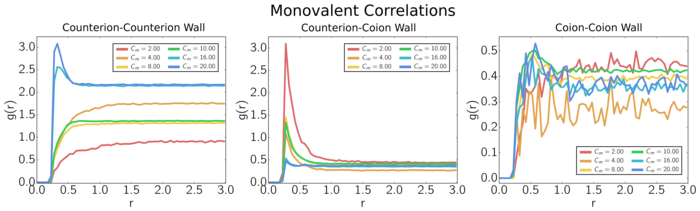}
\caption{Ion Correlations near the Wall.  The RDF $g(r)$ for ion-ion correlations in proximity to the wall for the  monovalent case with $\sigma = -6$, and $C_m = 2,4,8,10,16,20$.}
\label{fig:mono_ion_ion_correlations_wall}	
\end{figure}

}

\section{Ion-Ion Correlation Analysis}
\label{sec:corr_analysis}

\begin{figure}[H]
\centering
\includegraphics[width=0.8\columnwidth]{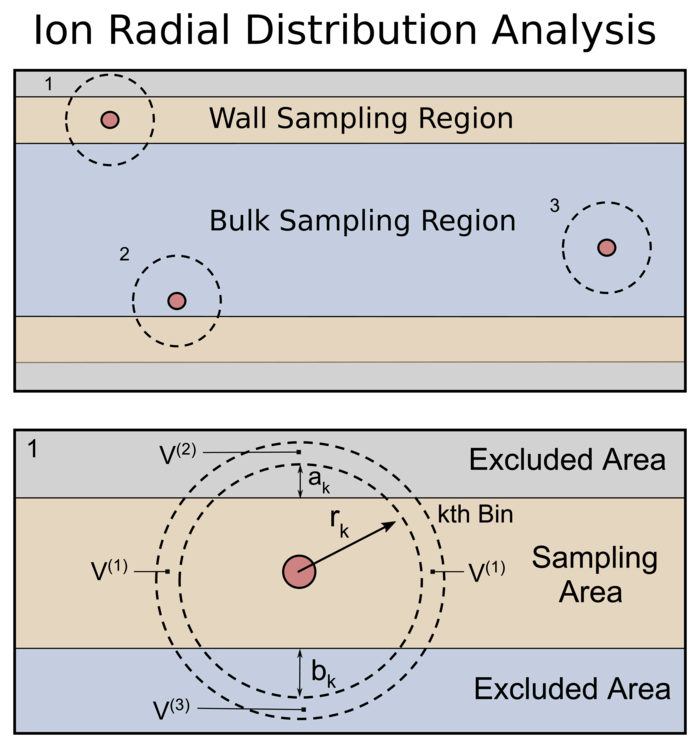}
\caption{Ion Radial Distribution Analysis.  To distinguish between the behaviors of the ions in the bulk vs near the channel wall in the condensed layer we perform regional sampling of a radial distribution function.  To avoid issues with ions excluded from the wall domain or within other sampling region we perform our radial distribution analysis $g(r)$ with probability conditioning on being within permissible regions.  We normalize the distribution at a given radius by the accessible volume $V^{(1)}$ of the ions which correspond to spherical caps.
}
\label{fig:radialDistFunc}
\end{figure}

We perform analysis of the radial distribution of the ions taking into account the proxmity of the ions to wall vs the bulk regions and by choosing carefully a normalization taking into account accessible regions of ions.  We split the channel into two sampling regions.  The first corresponds to the wall case when the base ion is within the distance $d < 1$nm from the channel wall.  The second is the bulk case  when the base ion is a distance $d > 1$nm from the channel wall.  In the confined channel geometry there are limited regions where ions are permitted given either the excluded volume of the wall or intrusion into the bulk or wall sampling region.  We handle this by a careful normalization by accessible volume to obtain a radial distribution function $g(r)$.  We give details below with a schematic of our approach in Figure~\ref{fig:radialDistFunc}.

For a bulk system the radial distribution function can be sampled for a base ion by counting the number of ions within a spherical shell at radius $r_k$ and thickness $\delta{r}$ to obtain the normalized distribution function $\bar{g}(r_k) = {H_k}/{V_k C_0}$.
The $V_k  =\frac{4\pi}{3}\left(R_k^3 - r_k^3\right)$ is the volume of the spherical shell of thickness $\delta{r}$, $R_k = r_k + \delta{r}$, $H_k$ is the histogram corresponding to the number of ions within the $k^{th}$ spherical shell, and $C_0$ is a normalizing constant typically chosen to correspond to the bulk concentration.  
To obtain a more spatially refined description of the ions taking into account excluded regions we define the radial distribution function as $g(r_k) = {\tilde{H}_k}/{\tilde{V}_k \tilde{C}_0}$ where $\tilde{H_k}$ for a given base ion is the histogram count for all permissible ions in the sampling region within the spherical shell of radius $r_k$ and thickness $\delta{r}$ and $\tilde{C}_0$ is a normalization based on the total concentration of ions.   To obtain a radial density we use the volume $\tilde{V}_k$ corresponding only to the part of the spherical shell that is within the permissible sampling region.  This can be computed using the geometry of spherical caps to obtain $\tilde{V}_k = V^{(1)} = V_k - V^{(2)} - V^{(3)}$ where $V^{(2)} = \frac{\pi}{3}\left(A_k^2\left(3R_k - A_k\right) - a_k^2\left(3r_k - a_k \right)  \right)$ and $V^{(3)} = \frac{\pi}{3}\left(B_k^2\left(3R_k - B_k\right) - b_k^2\left(3r_k - b_k \right)  \right)$ are the volumes associated with the shell of a spherical cap of thickness $\delta{r}$~\cite{bookHandbookEng2007}.  We denote by $A_k = a_k + \delta{r}$, $B_k = b_k + \delta{r}$, $R_k = r + \delta{r}$, see Figure~\ref{fig:radialDistFunc}.

Our radial distribution function can be thought of as the conditional probability function for a pair of ions occupying the sample sampling region.  Alternative methods have been considered in the literature such as sorting ions into $z$-slabs and sampling only in the $xy$-directions~\cite{bookReviewCompChemistry2015,Ryzsko2010}.  Both approaches provide very similar information and allow for distinguishing between the behaviors of ions in the bulk region and behaviors of ions in the condensed layer near to the walls.  

The approach we have introduced here allows for a unified observable that can transition from calculations involving sampling regions that are relatively narrow similar to z-slabs to intermediate and larger regions that yield results approaching the bulk radial distribution.  By use of this radial distribution function, we are able to obtain a refined understanding of how the ion correlations change when in regions in the bulk of the nanochannel versus when an ion occupies the condensed ion layer near to the wall which exhibits a quasi-two dimensional behavior.

\clearpage
\newpage

\end{multicols}

\end{document}